\documentclass[prd,notitlepage,nofootinbib,preprintnumbers,amssymb,superscriptaddress]{revtex4-1}
\usepackage{amsfonts,amssymb,amsmath,mathrsfs,graphicx,color,bm,ulem}
\usepackage{latexsym,subfigure,ascmac}
\usepackage{here}
\definecolor{ultramarine}{rgb}{0.07, 0.04, 0.56}
\definecolor{cadmiumgreen}{rgb}{0.0, 0.42, 0.24}
\definecolor{indigo(dye)}{rgb}{0.0, 0.25, 0.42}
\usepackage[linktocpage=true]{hyperref}
\hypersetup{
colorlinks=true,
citecolor=ultramarine,
linkcolor=cadmiumgreen,
urlcolor=indigo(dye),
}

\newcommand{\f}[2]{\frac{#1}{#2}}  
\newcommand{\mk}[1]{\left( #1 \right)}  
\newcommand{\kk}[1]{\left[ #1 \right]}  
\newcommand{\ck}[1]{\left\{ #1 \right\}}  
\newcommand{\be}{\begin{equation}}  
\newcommand{\ee}{\end{equation}}
\newcommand{\mR}{\mathcal{R}}
\newcommand{\mS}{\mathcal{S}}
\newcommand{\mO}{\mathcal{O}}

\newcommand{\mY}{\mathcal{Y}}
\newcommand{\mT}{\mathcal{T}}

\newcommand{\vp}{\varphi }

\newcommand{\df}{\dfrac}

\begin{document}

\title{
Exact solution for wave scattering from black holes: Formulation
}

\author{Hayato Motohashi}
\affiliation{Division of Liberal Arts, Kogakuin University, 2665-1 Nakano-machi, Hachioji, Tokyo, 192-0015, Japan}

\author{Sousuke Noda}
\affiliation{National Institute of Technology, Miyakonojo College, Miyakonojo 885-8567, Japan}
\affiliation{Division of Liberal Arts, Kogakuin University, 2665-1 Nakano-machi, Hachioji, Tokyo, 192-0015, Japan}
\affiliation{
  Center for Gravitation and Cosmology, College of Physical Science and Technology, 
  Yangzhou University, Yangzhou 225009, China}
\affiliation{School of Aeronautics and Astronautics, Shanghai Jiao Tong University, 200240 Shanghai, China}

\begin{abstract}
We establish an exact formulation for wave scattering of a massless field with spin and charge by a Kerr-Newman-de Sitter black hole.
Our formulation is based on the exact solution of the Teukolsky equation in terms of the local Heun function, and does not require any approximation.
It serves as simple exact formulae with arbitrary high precision, which realize fast calculation without restrictions on model parameters.
We highlight several applications including quasinormal modes, cross section, reflection/absorption rate, and Green function. 
\end{abstract}

\maketitle  


\section{Introduction}
\label{sec:intro}

Observational efforts to prove black holes (BHs) finally began to bear fruit in the last few years: direct detection of gravitational waves (GWs) emanating from a merger of binary BHs~\cite{Abbott:2016blz}, and electromagnetic observations with very-long-baseline interferometry (VLBI)~\cite{Akiyama:2019eap}.
The growing global network of ground-based GW interferometers and VLBI multi-wavelength observations at higher resolution in the near future will be powerful tools to unveil the nature of BHs or to test the Kerr hypothesis. 
By virtue of the uniqueness of the Kerr solution in General Relativity, they allow us unprecedented tests of gravity in the strong-field regime. 
Theoretical prediction of the propagation of fields with different spins on BH geometry is thus important.

While in the short-wavelength regime one can rely on the geometrical optics approximation, 
in the long-wavelength regime the approximation breaks down and one needs to take into account wave optics.
Most of the physically interesting cases of the wave equations on BH geometries can be solved by separation of variables.
The separability of the 
Klein-Gordon equation for the Kerr-Newman family with a cosmological constant was clarified by Carter~\cite{Carter:1968,Carter:1968rr,Carter:1968ks}.
This result, with the aid of the Newman-Penrose formalism~\cite{Newman:1961qr}, was generalized to higher-spin wave equations for the Kerr background~\cite{Teukolsky:1972my,Teukolsky:1973ha,Unruh:1973bda,Chandrasekhar:1976ap} 
and for the Kerr-de Sitter background~\cite{Khanal:1983vb,Chambers:1994ap}.
With the separated master equation known as the Teukolsky equation, one can investigate wave propagation, for which scattering analysis is a powerful approach~\cite{futterman_handler_matzner_1988}.
Observables in the wave optics have been commonly evaluated in the literature with certain approximations such as the WKB approximation.

In \cite{Suzuki:1998vy}, Suzuki, Takasugi, and Umetsu (STU) showed that 
both angular and radial parts of the Teukolsky equations for a massless field with spin and charge on 
the Kerr-Newman-de Sitter (KNdS) spacetime can be transformed into the 
Heun equation.\footnote{For the asymptotically flat spacetime, the Teukolsky equations can be transformed into the confluent Heun equation~\cite{Marcilhacy1983,Blandin1983,Gal'tsov1989,Mano:1996vt,Mano:1996mf,Mano:1996gn,Fiziev:2005ki,Borissov:2009bj,Fiziev:2009wn,Fiziev:2011mm,Bezerra:2013iha,Vieira:2016ubt}.  
See also \cite{Vieira:2014waa,Kraniotis:2016maw,Kraniotis:2018zmh,Hui:2019aqm,Bamber:2020bpu,Dariescu:2021zve} for analyses of massive fields in the context of the Heun equations.}
This result was further generalized to Petrov type-D vacuum backgrounds with a cosmological constant~\cite{Batic:2007it}. 
The Heun equation is a second-order linear homogeneous ordinary differential equation with four regular singular points~\cite{Heun1889,ronveaux1995heun,slavianov2000special,Maier_2006,Hortacsu:2011rr}.
There are several types of exact solutions for the Heun equation, depending on the analyticity around singular points.
Among them, in a series of works~\cite{Suzuki:1998vy,Suzuki:1999nn,Suzuki:1999pa}, STU adopted a series of the hypergeometric functions to construct an exact solution for the Teukolsky equation on the KNdS background, along the same lines as the Mano-Suzuki-Takasugi formalism~\cite{Mano:1996vt,Mano:1996mf,Mano:1996gn} for the asymptotically flat background.
They derived an exact formula for the absorption rate in terms of an infinite series.
Their formalism was also applied to the calculation of the quasinormal mode (QNM) frequencies for the Kerr-de Sitter black hole~\cite{Yoshida:2010zzb}, generalizing Leaver's method~\cite{Leaver:1985ax}.
On the other hand, in recent work~\cite{Hatsuda:2020sbn} on the Kerr-de Sitter background, Hatsuda employed 
a simpler exact solution known as the local Heun function or simply the local solution, 
and obtained a compact formula for the QNM frequencies with arbitrary high precision.

In this paper, we consider a massless test field with spin and charge on the KNdS background, and 
establish an exact formulation of the scattering problem using the local Heun function.  
The formulation based on the local Heun function is transparent and provides us with concise formulae for black hole physics such as the greybody factor and Green function. 
One can evaluate specific values of the local Heun function by using a modern technical computing system, Mathematica, which implemented the various Heun functions as built-in functions in the version 12.1 update in 2020.

The rest of the paper is organized as follows.
In \S\ref{sec:sol}, we transform the Teukolsky equation for a massless field on the KNdS background to the Heun equation, and provide the exact solution in terms of the local Heun function.  
In \S\ref{sec:scat}, we consider the boundary condition at the horizons, and obtain the connection coefficients, which allow us to solve the scattering problem with the exact solution.
In \S\ref{sec:app}, we highlight several applications of our formulation such as QNMs, S-matrix, cross section, reflection/absorption rate, and Green function.
\S\ref{sec:con} is devoted to the conclusion.

\section{Exact solution}
\label{sec:sol}

In this section, following \cite{Suzuki:1998vy,Hatsuda:2020sbn}, we present the exact solution of the Teukolsky equation on the KNdS background. 
After summarizing our notation of the KNdS metric in \S\ref{ssec:knds}, we review the transformation of the angular and radial parts of the Teukolsky equation into the Heun equation in \S\ref{ssec:tth}.  
In \S\ref{ssec:hl} we provide the exact solution in terms of the local Heun function.
We consider the boundary condition of the angular solution in \S\ref{ssec:asol}, and deal with the radial solution in \S\ref{sec:scat}.

\subsection{Kerr-Newman-de Sitter spacetime}
\label{ssec:knds}

As a rotating and charged black hole solution in the presence of the cosmological constant, 
we consider the KNdS spacetime.
The KNdS metric in Boyer-Lindquist coordinates takes the following form
\be ds^2 = - \f{\Delta}{(1+\alpha)^2\rho^2} (dt-a\sin^2\theta d\varphi)^2
+ \rho^2 \mk{ \f{dr^2}{\Delta} + \f{d\theta^2}{1+\alpha \cos^2\theta} } 
+ \f{ (1+\alpha \cos^2\theta)\sin^2\theta }{(1+\alpha)^2\rho^2} [adt-(r^2+a^2) d\varphi]^2
, \ee 
where 
\be \label{Ddef} \Delta(r)=(r^2+a^2)\mk{1-\f{\Lambda}{3}r^2}-2Mr+Q^2, \quad 
\alpha = \f{\Lambda a^2}{3}, \quad \rho^2=r^2+a^2 \cos^2\theta
. \ee
Here, $\Lambda$ is the cosmological constant, and 
$M, aM$, and $Q$ are respectively the mass, angular momentum, and charge of the black hole.
The electromagnetic field caused by the charge of the black hole is given by 
\be A_\mu dx^\mu = -\f{Qr}{(1+\alpha)^2\rho^2}(dt-a\sin^2\theta d\varphi)  .  \ee
One can consider several limiting cases.
For instance, $Q=0$ reproduces the Kerr-de Sitter spacetime, whereas $Q=0$ and $a=0$ reproduce the Schwarzschild-de Sitter (SdS) spacetime.

Throughout the paper, we assume $\Lambda> 0$, and focus on the case where $\Delta(r)=0$ has four distinct real roots
under the condition~\cite{Belgiorno:2008xn}
\be \alpha < 7 - 4 \sqrt{3}, \quad M_{c,-} < M < M_{c,+}, \ee
where 
\be M_{c,\pm} = \f{(1-\alpha)^{3/2}}{3\sqrt{2\Lambda}} \sqrt{1\pm \gamma} (2\mp \gamma) ,\quad \gamma = \sqrt{1-\f{12(\alpha+\beta)}{(1-\alpha)^2}  } ,\quad \beta=\f{\Lambda Q^2}{3} . \ee
We denote the four roots of $\Delta(r)=0$ as $r_{\pm},r'_{\pm}$.
We can then factorize $\Delta(r)$ as
\be \label{Ddef2} \Delta(r) = -\f{\Lambda}{3}(r-r_-)(r-r_+)(r-r'_+)(r-r'_-). \ee
We set the ordering of the four roots as $r'_-<0\leq r_-<r_+<r'_+$,
where $r_-, r_+$, and $r'_+$ are the inner (Cauchy) horizon, outer (event) horizon, and cosmological horizon, respectively.
We are interested in the scattering problem in the range $r_+ \leq r \leq r'_+$.
Comparing \eqref{Ddef2} with \eqref{Ddef}, it holds that
\begin{align} \label{solcoe}
r'_- + r_- + r_+ + r'_+ &= 0. 
\end{align}
Note that so long as $\Delta(r)=0$ has four distinct roots, our arguments in \S\ref{sec:sol} apply to the asymptotically AdS geometry with $\Lambda<0$, except for the ordering of the four roots.

For the KNdS spacetime with $\Lambda M^2\ll 1$, we have 
\be  
r_\pm \simeq M\pm\sqrt{M^2-a^2-Q^2},\quad
r'_\pm \simeq \pm \sqrt{\f{3}{\Lambda}} . \ee
For the SdS case, if $0< \Lambda M^2<1/9$, there are four real roots, which can be expressed in a simple expression
\be
r'_- = -2M\, {\rm Re}(\xi), \quad 
r_- = 0,\quad 
r_+ = 2M\, {\rm Re}(e^{i\pi/3}\xi),\quad
r'_+ = 2M\, {\rm Re}(e^{-i\pi/3}\xi), 
\ee
where
\be
\xi=\mk{\f{1}{\Lambda M^2}}^{1/3}\mk{3+i\sqrt{\f{1}{\Lambda M^2}-9}}^{1/3}.
\ee
For $\Lambda M^2\ll 1$, 
\be
\xi\simeq \f{e^{i\pi/6}}{\sqrt{\Lambda M^2}} (1-i\sqrt{\Lambda M^2}) ,
\ee
and hence
\be 
r_+ \simeq 2M,\quad
r'_\pm \simeq \pm \sqrt{\f{3}{\Lambda}} . \ee
For instance, for the SdS with $\Lambda M^2=10^{-3}$, we have $r'_-/M =-55.75$, $r_+/M =2.005$, $r'_+/M =53.74$.

The tortoise coordinate $r_*$ is defined by 
\be \label{drstardef} dr_*=\f{(1+\alpha)(r^2+a^2)}{\Delta(r)}dr, \ee
or 
\be \label{rstar} r_*= \f{\ln|r-r_+|}{2\kappa(r_+)} + \f{\ln|r-r_+'|}{2\kappa(r_+')} + \f{\ln|r-r_-'|}{2\kappa(r_-')} + \f{\ln|r-r_-|}{2\kappa(r_-)} , \ee
where 
\be \kappa(r_h)=\f{\Delta'(r_h)}{2(1+\alpha)(r_h^2+a^2)}, \ee
yields the surface gravity at the horizons.

\subsection{Transformation of the Teukolsky equation into the Heun equation}
\label{ssec:tth}

We consider the propagation of a massless field with spin $s$ and charge $e$ on the KNdS background.
In terms of the Newman-Penrose formalism, the master variables $\psi_s$ are given by 
\be 
\psi_s = 
\begin{cases}
\Psi_0 ~~{\rm or}~~ \rho^{-4}\Psi_4, & (s=2 ~~{\rm or}~~ -2), \\
\Phi_0 ~~{\rm or}~~ \rho^{-2}\Phi_2, & (s=1 ~~{\rm or}~~ -1), \\
\chi_0 ~~{\rm or}~~ \rho^{-1}\chi_1, & (s=\f{1}{2} ~~{\rm or}~~ -\f{1}{2}), \\
\phi & (s=0),
\end{cases}
\ee
where each case corresponds to the gravitational, electromagnetic, Dirac, and scalar field, respectively.
Note that $s=0$ corresponds to a conformally coupled massless scalar field, whose equation of motion is given by $(\Box-R/6)\phi=0$, 
where $\Box\phi=\f{1}{\sqrt{-g}}D_\nu(\sqrt{-g} g^{\mu\nu}D_\mu\phi)$ with $D_\mu=\partial_\mu-ieA_\mu$.

The Teukolsky equations for spin $0, \f{1}{2}, 1, \f{3}{2}, 2$ fields on the Kerr-de Sitter background and those for spin $0,\f{1}{2}$ fields on the KNdS are separable and take the unified form~\cite{Suzuki:1999pa}.
With 
\be \psi_s = R_s(r) S_s(\theta) e^{-i\omega t}e^{im\varphi} , \ee
and the separation constant $\lambda$,
the angular and radial parts of the Teukolsky equation are given by\footnote{There is a typo in the angular equation (3.1) in \cite{Suzuki:1999pa}: The second last term $-2m(1+\alpha)\xi$ in the second line should be $+2m(1+\alpha)\xi$.  With the corrected sign, Eq.~(3.1) is consistent with Eq.~(2.7) and their definition of $A_3$.  In \eqref{angTeu}, we rewrite Eq.~(3.1) in a more compact form, reflecting the correction.  
Equation~\eqref{angTeu} is also consistent with Eq.~(2.4) in \cite{Hatsuda:2020sbn}. 
}
\begin{align} 
\label{angTeu}
&\Biggl[ \f{d}{dx}(1+\alpha x^2)(1-x^2)\f{d}{dx} +\lambda -s(1-\alpha) -2\alpha x^2 \notag\\
&~~+ \f{4 s x (1 + \alpha) [m \alpha - c (1 + \alpha)]}{1 + \alpha x^2}
-\f{(1 + \alpha)^2 [m + s x - (1 - x^2) c]^2}{(1 + \alpha x^2) (1 - x^2)} \Biggr] S_s (x) = 0, \\
\label{radTeu}
&\Biggl[ \Delta^{-s}\f{d}{dr}\Delta^{s+1}\f{d}{dr} 
+ \f{J^2-isJ\Delta'}{\Delta} 
+2 i s J' 
- \f{2\alpha}{a^2}(s+1)(2s+1)r^2
+2s(1-\alpha)-\lambda \Biggr] R_s (r) = 0 .
\end{align}
where $x=\cos \theta$, $c=a\omega$, $\Delta'=d\Delta/dr$, 
and 
\begin{align}
J(r)&=(1+\alpha) K - eQr , \\
K(r)&=\omega(r^2+a^2)-am . 
\end{align}

It was clarified in \cite{Suzuki:1998vy} that the angular and radial Teukolsky equations on the KNdS background can be transformed into Heun equations, 
and the exact solution was constructed in terms of a series of hypergeometric functions.
Regarding the transformation from the Teukolsky equation to the Heun equation, there are $4!=24$ independent transformations depending on how to map the four regular singular points.
For the angular part we follow the transformation adopted in \cite{Suzuki:1998vy}, whereas for the radial part we follow the transformation adopted in \cite{Hatsuda:2020sbn} for the Kerr-de Sitter background, so that our parameter regions of interest, $-1\leq x\leq 1$ or $r_+\leq r\leq r'_+$ are mapped to $0\leq z\leq 1$, where $z$ is the independent variable after the transformation.

Further, it was shown in \cite{Batic:2007it} that, for massless field on Petrov type-D vacuum backgrounds with a cosmological constant, the separated Teukolsky equations can be transformed into Heun equations.
While we focus on \eqref{angTeu} and \eqref{radTeu} on the KNdS background, our analysis can be straightforwardly generalized to such a case.

\subsubsection{Angular part}
\label{ssec:ang}

Let us begin with the angular part~\eqref{angTeu} of the Teukolsky equation.
Since \eqref{angTeu} does not depend on the charge $Q$, the argument on the angular part remains the same regardless of the charge.
Note also that the cosmological constant $\Lambda$ enters the equation only via $\alpha=\Lambda a^2/3$. 
Therefore, for the nonrotating limit $a\to 0$, for which $\alpha$ (and $c$) vanishes, the equation does not depend on $\Lambda$.

For simpler geometries, the angular Teukolsky equation~\eqref{angTeu} allows a simple exact solution. 
For nonrotating black hole, i.e., Schwarzschild(-de Sitter) or Reissner-Nordstr\"om(-de Sitter) black hole, the exact solution is known as the spin-weighted spherical harmonics, $S_s(\theta)e^{im\varphi}={}_sY_{\ell m}(\theta,\varphi)$, with the eigenvalue $\lambda=\ell(\ell+1)-s(s-1)$.
The explicit form is given by
\be \label{sYlm}
_{s}Y_{\ell m}(\theta, \vp)=(-1)^m\sqrt{\df{(\ell+m)!(\ell-m)!(2\ell+1)}{4\pi(\ell+s)!(\ell-s)!}}\sin^{2\ell}\left(\df{\theta}{2}\right)
\sum_{p=0}^{\ell-s}\dbinom{\ell-s}{p}\dbinom{\ell+s}{p+s-m}(-1)^{\ell-p-s}e^{im\vp}\cot^{2p+s-m}\left(\df{\theta}{2}\right).
\ee
For the Kerr or Kerr-Newman geometry, the exact solution is denoted as the spin-weighted spheroidal function.
No analytic expression for the eigenvalue $\lambda$ is known in this case.

For the more general case of a rotating black hole in the presence of the cosmological constant, the spin-weighted spheroidal function is not the analytic solution.
However, we can still derive the exact solution since the angular equation~\eqref{angTeu} can be transformed into the Heun equation.
With a nonzero cosmological constant, the angular Teukolsky equation~\eqref{angTeu} has four regular singular points at $x=\pm 1, \pm i/\sqrt{\alpha}$ after removing a removable singularity at $x=\infty$.
We transform the independent and dependent variables as
\begin{align}
\label{zadef} z&= \f{(1-i/\sqrt{\alpha})(x+1)}{2(x-i/\sqrt{\alpha})},\\
\label{Sdef} S_s(x)&=z^{A_1}(z-1)^{A_2}(z-z_a)^{A_3}(z-z_\infty) y^{\rm (a)}_{s}(z) ,
\end{align}
to map the four regular singular points $(-1,1,-i/\sqrt{\alpha},i/\sqrt{\alpha})$ to $(0,1,z_a,\infty)$.
Here, the superscript ${\rm (a)}$ denotes the angular part.
Note that the boundaries $x=-1,1$ are now mapped to $z=0,1$, respectively.
Here, we denote $z_\infty=z|_{x\to \infty}$ and $z_a=z|_{x\to -i/\sqrt{\alpha}}$, namely, 
\be z_\infty=\f{1-i/\sqrt{\alpha}}{2}, \quad
z_a = -\f{(1-i/\sqrt{\alpha})^2}{4i/\sqrt{\alpha}}, \ee
and define
\be \label{defA}
A_1=\df{m-s}{2},\quad  A_2=-\df{m+s}{2},\quad  
A_3=\df{1}{2}\left[ s + i\mk{ \df{1+\alpha}{\sqrt{\alpha}}c -m \sqrt{\alpha} } \right], \quad 
A_4=\df{1}{2}\left[ s - i\mk{ \df{1+\alpha}{\sqrt{\alpha}}c -m \sqrt{\alpha} } \right],
\ee
which satisfy an identity
\be \label{Aid} A_1+A_2+A_3+A_4=0. \ee
The transformations~\eqref{zadef} and \eqref{Sdef} allow us to rewrite the angular equation~\eqref{angTeu} as 
\be \label{yangTeu} \f{d^2y^{\rm (a)}_{s}}{dz^2} + \mk{ \f{2A_1+1}{z} + \f{2A_2+1}{z-1} + \f{2A_3+1}{z-z_a} } \f{dy^{\rm (a)}_{s}}{dz} + \f{\rho_+\rho_-z+u}{z(z-1)(z-z_a)}y^{\rm (a)}_{s} = 0, \ee
where
\begin{align}
\rho_+ = 1,\quad 
\rho_- =1-2 A_4,\quad
u =-\left[\df{i\lambda}{4\sqrt{\alpha}}+\df{1}{2}+A_1 +\left(m+\df{1}{2}\right)(A_3 -A_4)\right].
\end{align}
Equation~\eqref{yangTeu} is nothing but the Heun equation, at which we shall take a closer look in \S\ref{ssec:hl}.

\subsubsection{Radial part}
\label{ssec:rad}

Next, we proceed to the radial part of the Teukolsky equation~\eqref{radTeu}.
The equation has four regular singular points at $r=r_\pm, r'_\pm$ after removing a removable singularity at $r=\infty$.
We transform the independent and dependent variables as
\begin{align} 
\label{zrdef} z&=\f{r'_+-r_-}{r'_+-r_+} \f{r-r_+}{r-r_-}, \\
\label{Rdef} R_s(r)&=z^{B_1} (z - 1)^{B_2} (z - z_r)^{B_3} (z - z_\infty)^{2 s + 1} y^{\rm (r)}_s(z)
\end{align}
to map the four regular singular points $(r_+,r'_+,r'_-,r_-)$ to $(0,1,z_r,\infty)$.
Here, the superscript ${\rm (r)}$ denotes the radial part.
To avoid notational complexity, here we use $z$ to denote the independent variable as in the angular part, but no confusion should occur as the arguments on the angular and radial parts are independent of each other.
Note that the black hole horizon $r=r_+$ and the cosmological horizon $r=r'_+$ are now mapped to $z=0,1$, respectively.
Therefore, again, the parameter range that we are interested in is $0\leq z\leq 1$.
Here we denote $z_\infty=z|_{r\to \infty}$ and $z_r=z|_{r\to r'_-}$, namely, 
\be z_\infty=\f{r'_+-r_-}{r'_+-r_+}, \quad
z_r = z_\infty\f{r'_--r_+}{r'_--r_-}, \ee
both of which are larger than unity.
Also, we define a purely imaginary function
\be B(r) = \f{iJ(r)}{\Delta'(r)} , \ee
and denote
\be B_1=B(r_+), \quad
B_2=B(r'_+), \quad 
B_3=B(r'_-), \quad 
B_4=B(r_-), \ee
which satisfy an identity
\be \label{Bid} B_1+B_2+B_3+B_4=0. \ee

With the transformations~\eqref{zrdef} and \eqref{Rdef} and the identities~\eqref{solcoe} and \eqref{Bid}, the radial Teukolsky equation~\eqref{radTeu} can be rewritten as
\be \label{yradTeu} \f{d^2y^{\rm (r)}_{s}}{dz^2} + \mk{ \f{2B_1+s+1}{z} + \f{2B_2+s+1}{z-1} + \f{2B_3+s+1}{z-z_r} } \f{dy^{\rm (r)}_{s}}{dz} + \f{\sigma_+\sigma_- z+v}{z(z-1)(z-z_r)}y^{\rm (r)}_{s} = 0, \ee
where
\begin{align}
\sigma_+&= 2s+1,\\
\sigma_-&= s+1-2B_4,\\ 
v&= \f{ \lambda - 2 s (1 - \alpha) - \f{\Lambda}{3} (s+1)(2s+1)(r_+ r_- + r'_+ r'_-) }{\f{\Lambda}{3} (r_- - r'_-) (r_+ - r'_+) } \notag\\
&~~~- \f{ i (2s+1) [ 2 (1 + \alpha) \{ \omega (r_+ r_- + a^2) - a m \} - e Q (r_+ + r_-) ] }{\f{\Lambda}{3} (r_- - r'_-) (r_- - r_+) (r_+ - r'_+)} . 
\end{align}
These expressions are much simpler than those in \cite{Suzuki:1998vy} and a natural generalization of those in \cite{Hatsuda:2020sbn} for $Q=0$.

\subsection{Local Heun function}
\label{ssec:hl}

In \S\ref{ssec:ang} and \S\ref{ssec:rad}, we see that we can transform the angular and radial Teukolsky equations into \eqref{yangTeu} and \eqref{yradTeu} respectively, which are the same type of differential equation, as pointed out first in \cite{Suzuki:1998vy}.
This type of differential equation, i.e., the second-order Fuchsian equation with four regular singular points on the Riemann sphere, is known as the Heun equation~\cite{ronveaux1995heun,slavianov2000special,Maier_2006}, which is given by
\be \label{Heun} \f{d^2y}{dz^2} + \mk{ \f{\gamma}{z} + \f{\delta}{z-1} + \f{\epsilon}{z-a} } \f{dy}{dz} + \f{\alpha\beta z-q}{z(z-1)(z-a)} y = 0 , \ee
with the condition 
\be \label{Hcond} \gamma+\delta+\epsilon=\alpha+\beta+1, \quad a\ne 0,1. \ee
The Heun equation has six independent parameters.
$a$ is called a singularity parameter,  
$\alpha, \beta, \gamma, \delta$ (and $\epsilon$) are called exponent parameters, 
and $q$ is called an accessory parameter.
In \S\ref{ssec:hl} only, we use $\alpha, \beta, \gamma, a$ to denote the parameters of the Heun equation, rather than the parameters for the KNdS geometry.

The angular and radial Teukolsky equations in the forms \eqref{yangTeu} and \eqref{yradTeu} are nothing but the Heun equation~\eqref{Heun} with 
\be \label{param-ang} a=z_a, \quad 
q=-u,\quad
\alpha=\rho_+,\quad
\beta=\rho_-,\quad
\gamma=2A_1+1,\quad
\delta=2A_2+1,\quad
\epsilon=2A_3+1, \ee
and 
\be \label{param-rad} a=z_r, \quad 
q=-v,\quad
\alpha=\sigma_+,\quad
\beta=\sigma_-,\quad
\gamma=2B_1+s+1,\quad
\delta=2B_2+s+1,\quad
\epsilon=2B_3+s+1, \ee
respectively.
Note that the conditions~\eqref{Hcond} are satisfied by virtue of the identities~\eqref{Aid} and \eqref{Bid}.

The Heun equation has four regular singular points at $z=0,1,a,\infty$.
At the vicinity of each regular singular point, we can construct two linearly independent local solutions, or Frobenius solutions.
Following the standard notation, we denote the local Heun function $Hl(a, q;\alpha, \beta, \gamma, \delta;z)$ as the canonical local solution of the Heun equation at $z=0$, namely, 
\be \label{Hldef} Hl(a,q;\alpha, \beta,\gamma,\delta;z) = \sum^\infty_{k=0} c_kz^k, \ee
where the coefficients $c_k$ are defined by the three-term recurrence relation
\begin{align} \label{ckre}
&c_{-1}=0 \quad c_0=1, \notag\\
&(k+1)(k+\gamma)ac_{k+1} -\ck{ k[(k+\gamma+\delta-1)a+(k+\gamma+\epsilon-1)]+q }c_k
+(k+\alpha-1)(k+\beta-1)c_{k-1}=0.
\end{align}
The local Heun function~\eqref{Hldef} converges for $|z|<{\rm min}(1,|a|)$.
Therefore the maximum of the radius of convergence is unity for $|a|>1$. 
However, the local Heun function $Hl$ can be analytic at $z=0,1$ for some discrete values $q=q_m$ ($m=0,1,2,\cdots$).
In this case the function is called the Heun function and is denoted by $Hf$.
Further, it can be analytic at $z=0,1,a$ with $\alpha=-n$ ($n=0,1,2,\cdots$) and $q=q_m$ ($m=0,1,2,\cdots,n$). 
In this case the function becomes polynomial and is called the Heun polynomial $Hp$.
In this paper, we only use the local Heun function~\eqref{Hldef}.

The local Heun functions at $z=0,1$ are of special interest to us in discussing scattering from black holes.
Two local Heun functions at $z=0$ are given by 
\begin{align}
\label{y01def} y_{01}(z) &= Hl(a, q;\alpha, \beta, \gamma, \delta;z), \\
\label{y02def} y_{02}(z) &= z^{1-\gamma}Hl(a, (a\delta+\epsilon)(1-\gamma)+q;\alpha+1-\gamma, \beta+1-\gamma, 2-\gamma, \delta;z),
\end{align}
and two local Heun functions at $z=1$ are given by
\begin{align}
\label{y11def} y_{11}(z) &= Hl(1-a, \alpha\beta-q;\alpha, \beta, \delta, \gamma;1-z), \\
\label{y12def} y_{12}(z) &= (1-z)^{1-\delta}Hl(1-a, ((1-a)\gamma+\epsilon)(1-\delta)+\alpha\beta-q;\alpha+1-\delta, \beta+1-\delta, 2-\delta, \gamma;1-z).
\end{align}
The asymptotic behavior of the exact solutions~\eqref{y01def}--\eqref{y12def} is determined by the characteristic exponents 
\begin{align}
\label{asymy1} y_{01}(z)&= 1 + \mO(z), &  y_{02}(z)&=  z^{1-\gamma}[1+\mO(z)], & (z&\to 0) , \\
\label{asymy2} y_{11}(z)&= 1 + \mO(1-z), &  y_{12}(z)&=  (1-z)^{1-\delta}[1+\mO(1-z)], & (z&\to 1) .
\end{align}

The local Heun functions at $z=0$ are related to the local Heun functions at $z=1$ via linear combinations
\begin{align}
y_{01}(z)&= C_{11}y_{11}(z)+C_{12}y_{12}(z) , \label{y01rel} \\
y_{02}(z)&= C_{21}y_{11}(z)+C_{22}y_{12}(z) . \label{y02rel} 
\end{align}
The connection coefficients are formally given by the ratio of the Wronskians as
\be \label{C11} C_{11}=\f{W_z[y_{01}, y_{12}]}{W_z[y_{11}, y_{12}]}, \quad
C_{12}=\f{W_z[y_{01}, y_{11}]}{W_z[y_{12}, y_{11}]}, \quad 
C_{21}=\f{W_z[y_{02}, y_{12}]}{W_z[y_{11}, y_{12}]}, \quad 
C_{22}=\f{W_z[y_{02}, y_{11}]}{W_z[y_{12}, y_{11}]},
\ee
where $W_z[u,v] = u \f{dv}{dz} - \f{du}{dz} v$.
Note that from \eqref{yradTeu} it holds that, for linearly independent solutions $y_a,y_b$, 
\be z^{\gamma}(z-1)^{\delta}(z-z_r)^{\epsilon}W_z[y_a,y_b] = {\rm const}. \ee
Therefore, while the Wronskian itself is not constant, the ratio between two Wronskians is constant.

Conversely, the local Heun functions at $z=1$ can be expressed as
\begin{align}
y_{11}(z) = D_{11}y_{01}(z)+D_{12}y_{02}(z), \label{y11rel} \\ 
y_{12}(z) = D_{21}y_{01}(z)+D_{22}y_{02}(z), \label{y12rel}  
\end{align}
where 
\be 
\begin{pmatrix}
D_{11} & D_{12} \\
D_{21} & D_{22}
\end{pmatrix}
=
\begin{pmatrix}
C_{11} & C_{12} \\
C_{21} & C_{22}
\end{pmatrix}^{-1}
=
\f{W_z[y_{11}, y_{12}]}{W_z[y_{01}, y_{02}]}
\begin{pmatrix} \label{CDrel}
C_{22} & -C_{12} \\
-C_{21} & C_{11}
\end{pmatrix}
\ee
namely,
\be \label{D11} D_{11}=\f{W_z[y_{11},y_{02}]}{W_z[y_{01},y_{02}]}, \quad
D_{12}=\f{W_z[y_{11},y_{01}]}{W_z[y_{02},y_{01}]}, \quad 
D_{21}=\f{W_z[y_{12},y_{02}]}{W_z[y_{01},y_{02}]}, \quad
D_{22}=\f{W_z[y_{12},y_{01}]}{W_z[y_{02},y_{01}]}.\ee

While the connection coefficients can be formally written down analytically~\cite{Dekar1998}, this approach requires the evaluation of the local Heun function on the maximum convergence radius, and in general it is not clear whether it is convergent~\cite{Hortacsu:2020bee}.
Even if it is convergent, it typically requires the analytic continuation of the local Heun function, which has a high computational cost.
The expressions \eqref{C11} or \eqref{D11} are more practical.
To obtain the connection coefficients $C_{ij}$ or $D_{ij}$, one can evaluate the right-hand sides of \eqref{C11} or \eqref{D11} at any $z$ within the overlapping region of the two disks of convergence.
The advantage of this formulation is that the scattering problem is defined between $z=0$ and $1$ and the calculation remains within the circle of convergence of local Heun functions at $z=0$ and $1$.
This situation should be compared with the case where one needs a calculation outside the circle of convergence, for which one needs analytic continuation or other types of exact solutions of the Heun equation valid for a wider range, such as hypergeometric function series. 
In our case, we can calculate the connection coefficients at some point between $z=0$ and $1$ without analytic continuation.
We shall see in \S\ref{sec:scat} that the connection coefficients play a central role for the scattering problem.

For the specific calculations in the present paper, we use the built-in function {\tt HeunG} implemented in Mathematica 12.1 or later, which yields the local Heun function $Hl$~\eqref{Hldef} inside the circle of convergence, whereas it gives an analytic continuation of $Hl$ outside the circle of convergence.
The analytic continuation typically takes more computational time, and sometimes causes a multi-value issue.
For the radial Teukolsky equation, since $a=z_r>1$ holds, the radius of convergence for the local Heun functions~\eqref{y01def} and \eqref{y02def} at $z=0$ is unity.
Therefore, there always exists an overlapping region of the two disks of convergence for the local Heun functions at $z=0$ and $z=1$, where we can use both local Heun functions without analytic continuation.
The general solution of the radial Teukolsky equation~\eqref{yradTeu} can thus be written as a linear combination of $y^{\rm (r)}_{01,s},y^{\rm (r)}_{02,s}$ or $y^{\rm (r)}_{11,s},y^{\rm (r)}_{12,s}$.
Here, $y^{\rm (r)}_{Ii,s}$ denotes the radial exact solution, i.e., the exact solution $y_{Ii}$ with the parameter set~\eqref{param-rad} for $I=0,1$ and $i=1,2$.  
We define the angular exact solution $y^{\rm (a)}_{Ii,s}$ in the same manner with the parameter set~\eqref{param-ang}.  
For the scattering problem, we shall focus on two specific radial solutions imposing a certain set of boundary conditions, which we shall discuss in \S\ref{sec:scat}.
We shall also see that both local Heun functions are useful to see the asymptotic behavior close to the black hole horizon or cosmological horizon.

\subsection{Angular solution}
\label{ssec:asol}

Before proceeding to the scattering problem with the radial solution in \S\ref{sec:scat}, let us check the requirement on the regularity of the angular solution in terms of the exact solutions.  
Since the angular Teukolsky equation~\eqref{angTeu} does not depend on the charge $Q$, we can directly apply the argument of the angular part in \cite{Hatsuda:2020sbn} for the Kerr-de Sitter case.
From \eqref{asymy1} and \eqref{asymy2}, we see that the angular function $S_{Ii,s}=z^{A_1}(z-1)^{A_2}(z-z_a)^{A_3}(z-z_\infty) y^{\rm (a)}_{Ii,s}(z)$ satisfies
\begin{align}
\label{asymS1} S_{01,s}(x)&\propto (1+x)^{(m-s)/2}[1+\mO(1+x)], & S_{02,s}(z)&\propto (1+x)^{(s-m)/2}[1+\mO(1+x)], & (x&\to -1) , \\
\label{asymS2} S_{11,s}(x)&\propto (1-x)^{-(m+s)/2}[1+\mO(1-x)], & S_{12,s}(z)&\propto  (1-x)^{(m+s)/2}[1+\mO(1-x)], & (x&\to 1) .
\end{align} 
The general solution $S_s(x)$ is given by a linear combination of $S_{Ii,s}(x)$.
To make the angular solution regular at $x=\pm 1$, we should respectively choose $S_{01,s}(x)$ or $S_{02,s}(x)$ for $s-m\lesseqgtr 0$, and $S_{11,s}(x)$ or $S_{12,s}(x)$ for $m+s\lesseqgtr 0$.
For $S_s(x)$ to satisfy both regularities at $x=\pm 1$, we require linear dependence of the exact solutions, namely,
\be \label{exactlambda}
W_z[y^{\rm (a)}_{0i,s},y^{\rm (a)}_{1j,s}]=0,\quad
i=\begin{cases}
1, & (m-s\geq 0) , \\
2, & (m-s< 0) ,
\end{cases} \quad
j=\begin{cases}
1, & (m+s\leq 0) , \\
2, & (m+s> 0) ,
\end{cases} 
\ee
For a nonrotating black hole with $a/M=0$, this equation is satisfied by the eigenvalue $\lambda=\ell(\ell+1)-s(s-1)$.
For a rotating black hole, this equation depends on $\lambda$ and $\omega$ implicitly. 
For a fixed frequency $\omega$, this condition determines $\lambda$, which we can obtain by using a root-finding algorithm.
On the other hand, to obtain the QNM frequencies, we should solve \eqref{exactlambda} and a boundary condition on the radial solution to obtain $\lambda$ and $\omega$ simultaneously, as we shall see in \S\ref{sec:scat} and \S\ref{sec:app}.
In either case, we need an initial input value sufficiently close to the roots.

\begin{figure}[H]
 \centering
 \includegraphics[width=0.96\linewidth]{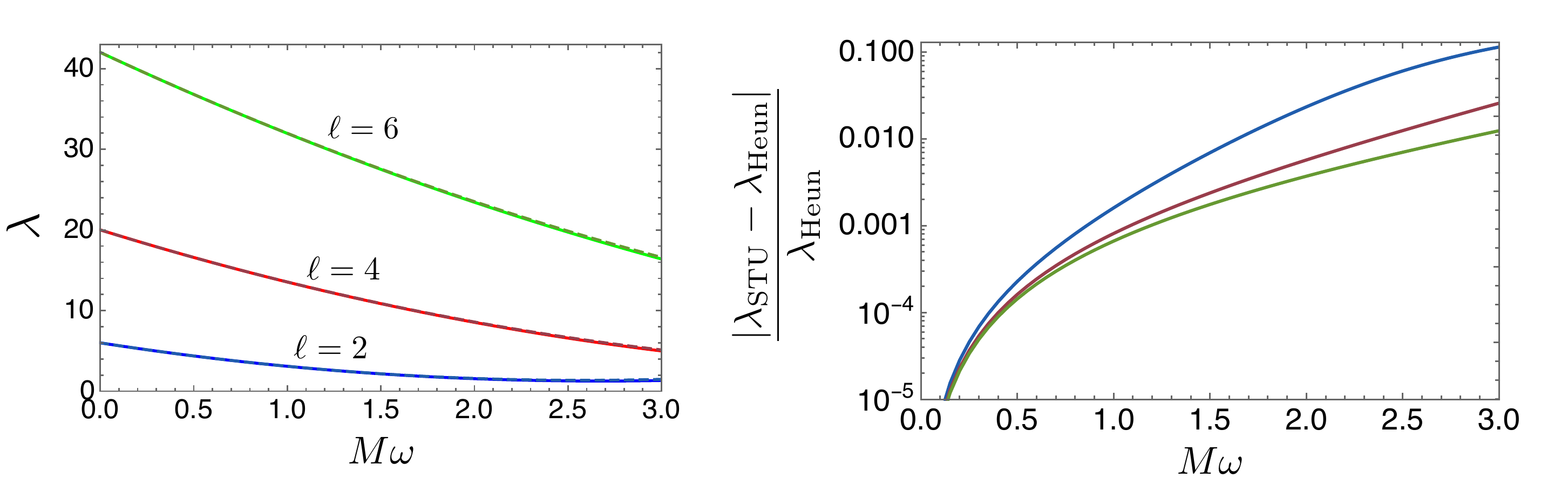}
 \caption{\footnotesize{Left: The eigenvalue $\lambda$ evaluated as the root of the exact formula~\eqref{exactlambda} (solid curves) and that obtained by the analytic expansion formula (4.18) in \cite{Suzuki:1998vy} for the scalar wave $(s=0)$ of $m=\ell$ and $\ell=2$~(blue), $4$~(red), $6$~(green) scattered by the Kerr-de Sitter black hole with $a/M=0.9$ and $\Lambda M^2=10^{-3}$. Right: The relative errors of the analytic expansion formula.}}\label{fig:lambdas}
\end{figure}

In Fig.~\ref{fig:lambdas}, we present the eigenvalue $\lambda$ for scalar waves on the Kerr-de Sitter background obtained by the above method.  
We compare our exact results with the analytic expansion formula given by Eq.~(4.18) in \cite{Suzuki:1998vy} for small $a\omega$ and $\Lambda a^2/3$. 
We denote these two results as $\lambda_\text{Heun}$ and $\lambda_\text{STU}$, respectively.
So long as one considers low-frequency waves scattered by a slowly rotating black hole with a small cosmological constant, the analytic expansion formula works well and the difference between $\lambda_\text{Heun}$ and $\lambda_\text{STU}$ is negligible.
To see its validity and limitation, we consider a rapidly rotating black hole $a/M=0.9$ with a small cosmological constant $\Lambda M^2=10^{-3}$.
In the left panel of Fig.~\ref{fig:lambdas}, $\lambda_\text{Heun}$ and $\lambda_\text{STU}$ are shown by solid and dashed curves, respectively, for $m=\ell$ and $\ell=2,4,6$.
For the calculation of $\lambda_\text{Heun}$, we pick up sampling points with the interval $\Delta(M\omega)=0.05$ for the range $0\leq M\omega \leq 3$. 
We take $\lambda=\lambda_{\text{STU}}$ as the initial input value for the root-finding algorithm {\tt FindRoot} in Mathematica, and set {\tt PrecisionGoal}~$\rightarrow 15$. 
For the algorithm to work well with this initial input, 
we need to set {\tt PrecisionGoal} larger than $12$.
To get the plots in Fig.~\ref{fig:lambdas}, we use {\tt ParallelTable} with 8 cores 
and get the list of data. 
The computation time for each curve is about $2.5$~sec.
In the right panel of Fig.~\ref{fig:lambdas}, we present the relative errors between $\lambda_\text{Heun}$ and $\lambda_\text{STU}$.
As expected, the relative error increases as the frequency increases.
In this setup, we see that for $M\omega\leq 1$ and $\ell\geq 2$, the error remains $\mO(10^{-1})$\%, so it is reasonable for this parameter range to use the analytic expansion formula.  On the other hand, for low-multipole and high-frequency waves, the error of the analytic expansion formula becomes large, and hence one should use the exact formula. 

As a test of the application range of the present method, we check the case where $\lambda=0$ is adopted as the 
initial input value. For this initial value, the root-finding algorithm requires a longer computational time and larger value of {\tt PrecisionGoal} since the initial values for larger $\ell$ are far from the true value. For example, if we set {\tt PrecisionGoal} smaller than $15$ for $\ell=2$, the method does not work well with the initial input $\lambda=0$. 
In practice, one can also adopt the eigenvalue $\lambda=\ell(\ell+1)-s(s-1)$ for the nonrotating case ($a=0$) as a simpler initial input value than $\lambda_\text{STU}$, while in that case the computation time becomes about 1.5 times as long as the case with $\lambda_\text{STU}$. 
However, the precision reaches, e.g., 20 digits so long as one requires the option {\tt PrecisionGoal}~$\rightarrow 20$.

\section{Scattering problem}
\label{sec:scat}

In this section we focus on the radial solution and provide the exact solution for the scattering problem.
In \S\ref{ssec:asym} we consider the asymptotic solution at the black hole and cosmological horizons, respectively.
We shall see that the asymptotic solutions correspond to in/outgoing waves and are consistent with the asymptotic form of the exact solution in terms of the local Heun function obtained in \S\ref{sec:sol}.
In \S\ref{ssec:scat} we exploit the asymptotic solution as the boundary condition, and write down the coefficients for in/outgoing waves in terms of the connection coefficients for the local Heun function.

For the following we omit the superscript ${\rm (r)}$ from the radial solution $y^{\rm (r)}_{Ii,s}$ for simplicity. 
Since we do not discuss the angular solution $y^{\rm (a)}_{Ii,s}$ below, no confusion should occur.

\subsection{Asymptotic behavior}
\label{ssec:asym}

We can obtain the boundary condition by considering the asymptotic behavior of the radial equation at the black hole and cosmological horizons, for which the Schr\"odinger form is useful.
We employ the tortoise coordinate $r_*$ defined in \eqref{drstardef} as an independent variable, and transform the dependent variable as
\be \label{def-Y} \mY_s=\Delta^{s/2}(r^2+a^2)^{1/2} R_s . \ee
We can then rewrite the radial Teukolsky equation~\eqref{radTeu} in the Schr\"odinger form 
\be \label{diffeq-Y} \mk{\f{d^2}{dr_*^2} + V_s } \mY_s = 0, \ee
with the potential
\begin{align} \label{Vs} 
&V_s(r)= \f{1}{(1+\alpha)^2(r^2+a^2)^2} \mk{J - \f{i s \Delta'}{2} }^2  \notag\\
&+ \f{\Delta}{(1+\alpha)^2(r^2+a^2)^2} \kk{ 2 i s \ck{2 (1 + \alpha) \omega r - e Q} 
- \f{2}{3} \Lambda r^2 (2 s^2 + 1) + 
s (1 - \alpha) - \lambda 
+ \f{2 r^2 - a^2}{(r^2+a^2)^2} \Delta 
- \f{r}{r^2+a^2} \Delta' }.
\end{align}

The potential depends on the spin only via $s^2$ and $is$, except $s (1 - \alpha) - \lambda$, apparently.
Actually, the combination $s (1 - \alpha) - \lambda$ is invariant under $s\to -s$~\cite{Suzuki:1999pa}.
Hence, the potential~\eqref{Vs} has a symmetry $V^*_{-s}(r)=V_s(r)$, where $z^*$ is a complex conjugate of $z$.
This implies that $\mY_s(r_*)$ and $\mY_{-s}^*(r_*)$ are two linearly independent solutions of the same differential equation~\eqref{diffeq-Y}.
Therefore, if $R_s=\Delta^{-s/2}(r^2+a^2)^{-1/2}\mY_{s}$ is a solution of the radial Teukolsky equation, $\Delta^{-s}R_{-s}^*=\Delta^{-s/2}(r^2+a^2)^{-1/2}\mY_{-s}^*$ is the solution linearly independent to $R_s$.

The potential asymptotically approaches a constant value
\be\label{Vasym} V_s(r)\to  -\f{\Delta'^2_h}{(1+\alpha)^2(r_h^2+a^2)^2} \mk{B_h+\f{s}{2} }^2, \quad (r\to r_h) , \ee
where we denote $r_1=r_+$, $r_2=r_+'$, and 
$f_h=f(r_h)$ for $h=1$, $2$.
Consequently, the asymptotic behavior of two independent solutions is given by
\be \label{Yasym}
\mY_s(r_*) \to 
\exp\kk{ \pm \f{\Delta'_h}{(1+\alpha)(r_h^2+a^2)} \mk{B_h+\f{s}{2} } r_*}, \quad (r\to r_h) .
\ee
From \eqref{rstar}, at the vicinity of the horizon $r_+$ or $r'_+$, the tortoise coordinate behaves as 
\be \label{asymtort} r_* \to \f{(1+\alpha)(r_h^2+a^2)}{\Delta'_h}\ln |r-r_h|,\quad (r\to r_h) .  \ee
Using \eqref{asymtort}, we obtain 
\be \label{asymtort2} \exp\kk{ \f{\Delta'_h}{(1+\alpha)(r_h^2+a^2)} r_*} \to | r-r_h | \simeq \left| \f{\Delta(r)}{\Delta'_h} \right|, \quad (r\to r_h) . \ee
Plugging \eqref{asymtort2} into \eqref{Yasym} and multiplying $\Delta^{-s/2}(r^2+a^2)^{-1/2}$, we obtain the asymptotic solutions of the radial Teukolsky equation 
\be \label{Rasym} R_s (r)\to \Delta^{B_h} ~~\text{and}~~ \Delta^{-B_h-s}, \quad (r\to r_h), \ee
where we have omitted proportional constants.
One can check that for the SdS case the asymptotic solutions~\eqref{Rasym} are $e^{i\omega r_*}$ and $\Delta^{-s}e^{-i\omega r_*}$, respectively.

\subsection{Scattered waves}
\label{ssec:scat}

In general, the asymptotic behavior of a general solution $R_s(r)$ is given by a linear combination of the two asymptotic solutions~\eqref{Rasym}.
For the scattering problem, we focus on two independent solutions $R_{\rm in}(r)$ and $R_{\rm up}(r)$ that satisfy the following asymptotic behaviors~\cite{PhysRevD.10.1701} 
\begin{align}
\label{Rin-asym} R_{{\rm in},s}(r) &\to
\begin{cases}
\displaystyle C_s^{\rm (trans)} \Delta^{-B_1-s} , & (r\to r_+) ,\\
\displaystyle C_s^{\rm (ref)} \Delta^{B_2} + C_s^{\rm (inc)} \Delta^{-B_2-s} , & (r\to r'_+) ,
\end{cases}\\
\label{Rup-asym} R_{{\rm up},s}(r) &\to
\begin{cases}
\displaystyle D_s^{\rm (up)} \Delta^{B_1} + D_s^{\rm (ref)} \Delta^{-B_1-s} , & (r\to r_+) ,\\
\displaystyle D_s^{\rm (trans)} \Delta^{B_2} , & (r\to r'_+) .
\end{cases}
\end{align}
The physical meaning is transparent once combined with the time-dependent part $e^{-i\omega t}$.  
The ``in'' solution is defined by the boundary condition that there is no wave coming out from the black hole horizon.
On the other hand, the ``up'' solution is defined by the boundary condition that there is no incoming wave from the cosmological horizon.
Both boundary conditions are appropriate for the classical picture of the horizons.
Combined with two other solutions defined by 
\be \label{Rod-asym} R_{{\rm out},s}=\Delta^{-s}R_{{\rm in},-s}^*, \quad R_{{\rm down},s}=\Delta^{-s}R_{{\rm up},-s}^*, \ee 
any two solutions among the four solutions~\eqref{Rin-asym}--\eqref{Rod-asym} are linearly independent solutions for the same radial Teukolsky equation.
For the scattering problem, we mainly use $R_{{\rm in},s}$ and $R_{{\rm up},s}$.

In the definition of $R_{{\rm in},s}$ in \eqref{Rin-asym} and $R_{{\rm up},s}$ in \eqref{Rup-asym} there are six coefficients. 
Not all the coefficients are independent.
Clearly, one can omit the overall factors as the degrees of freedom for the normalization, but here we keep them for later convenience.
On the other hand, we can derive relations between coefficients for $R_{{\rm in},s}$ and $R_{{\rm up},s}$ as follows.
From \eqref{radTeu}, for a set of two linearly independent solutions $R_1,R_2$, it holds that 
\be \label{RdRconst} \Delta^{s+1} W_r [R_1, R_2] = {\rm const,} \ee
where $W_r [R_1, R_2] = R_1 \f{dR_2}{dr} - \f{dR_1}{dr} R_2$.
Plugging in $(R_1,R_2)=(R_{{\rm in},s},R_{{\rm up},s})$ and $(R_{{\rm out},s},R_{{\rm up},s})$, we obtain 
\begin{align} 
\label{CsDsconst1}  \f{D_s^{\rm (trans)}}{D_s^{\rm (up)}}  
&= F_s \f{C_s^{\rm (trans)}}{C_s^{\rm (inc)}}   , \\
\label{CsDsconst2}  \f{D_s^{\rm (trans)}}{D_s^{\rm (ref)}}  
&= -F_s \f{C_{-s}^{\rm (trans)*}}{C_{-s}^{\rm (ref)*}} ,
\end{align}
where
\be \label{Fs} F_s=\f{\Delta'(r_+) (2B_1+s)}{\Delta'(r'_+) (2B_2+s)}. \ee
Note that $F_{-s}^*=F_s$ holds.
The ratios between the coefficients~$C_s, D_s$ determine the scattering problem and yield the S-matrix, reflection/transmission rate, and so on. 
Our aim in this section is thus to write down the coefficients~$C_s, D_s$ using the exact solution in terms of the local Heun function given in \S\ref{sec:sol}.

As we shall see below, the asymptotic behavior suggests that $R_{{\rm in},s}(r), R_{{\rm up},s}(r)$ respectively corresponds to $y_{02,s}(z), y_{11,s}(z)$, namely,
\begin{align} 
\label{Rin-Heun} R_{{\rm in},s}(r) &= \begin{cases}
R_{02,s}(r), & (r\to r_+) ,\\
C_{21,s}R_{11,s}(r)+C_{22,s}R_{12,s}(r), & (r\to r'_+) ,
\end{cases}\\
\label{Rup-Heun} R_{{\rm up},s}(r) &= \begin{cases}
D_{11,s}R_{01,s}(r)+D_{12,s}R_{02,s}(r), & (r\to r_+) ,\\
R_{11,s}(r), & (r\to r'_+) ,
\end{cases}
\end{align} 
where each $R_{Ii,s}$ is defined by \eqref{Rdef} with the corresponding solution $y_{Ii,s}$, with $I=0,1$ and $i=1,2$.
Note that here we are not using any approximation but using the exact relations~\eqref{y02rel} and \eqref{y11rel}.
$R_{{\rm in},s}, R_{{\rm up},s}$ are given exactly by the local Heun functions at $z=0$ and $z=1$, and each two expressions coincide with each other for the region where two disks of convergence overlap.

Using 
$r-r_h \simeq \Delta(r)/\Delta'_h$ for $r\to r_h$, 
we obtain
\begin{align} 
z &\simeq A \Delta(r), \quad\, (z\to 0;\, r\to r_+) ,\\
1-z &\simeq A' \Delta(r), \quad (z\to 1;\, r\to r'_+) ,
\end{align}
where
\begin{align} 
A= \f{z_\infty}{(r_+-r_-)\Delta'(r_+)}, \quad
A'=\f{z_\infty(r_+-r_-)}{-(r'_+-r_-)^2\Delta'(r'_+)}.
\end{align}
With these relations and the asymptotic expansions~\eqref{asymy1} and \eqref{asymy2}, we see that the solutions \eqref{Rin-Heun}, \eqref{Rup-Heun} indeed satisfy the boundary conditions given in \eqref{Rin-asym}, \eqref{Rup-asym}, respectively.

Hence, we can express the coefficients $C_s, D_s$ in \eqref{Rin-asym} and \eqref{Rup-asym} as 
\begin{align} 
C_s^{\rm (inc)} &= C_{22,s} (- 1)^{B_2} (1 - z_r)^{B_3} (1-z_\infty)^{2s+1} A'^{-B_2-s}, \label{eq:Cinc} \\
D_s^{\rm (up)} &= D_{11,s} (- 1)^{B_2} (- z_r)^{B_3} (- z_\infty)^{2s+1} A^{B_1}, \\
C_s^{\rm (ref)} &= C_{21,s} D_s^{\rm (trans)} \notag\\
&= C_{21,s} (- 1)^{B_2} (1 - z_r)^{B_3} (1-z_\infty)^{2s+1} A'^{B_2} , \label{eq:Cref} \\
D_s^{\rm (ref)} &= D_{12,s} C_s^{\rm (trans)} \notag\\
&= D_{12,s} (- 1)^{B_2} (- z_r)^{B_3} (- z_\infty)^{2s+1} A^{-B_1-s} . \label{eq:Ctrans} 
\end{align}
For the scattering problem, the ``squares'' of the coefficients are important; these take the following form without $A$ and $A'$:
\begin{align} 
\label{Cincsq} C_s^{\rm (inc)} C_{-s}^{\rm (inc)*} &= C_{22,s} C_{22,-s}^* e^{2i\pi(B_2+B_3)} (z_\infty-1)^{2} , \\
C_s^{\rm (ref)} C_{-s}^{\rm (ref)*} &= C_{21,s}C_{21,-s}^* e^{2i\pi(B_2+B_3)} (z_\infty-1)^{2}  ,\\
\label{Ctranssq} C_s^{\rm (trans)} C_{-s}^{\rm (trans)*} &= e^{2i\pi(B_2+B_3)} z_\infty^{2} , \\
D_s^{\rm (up)} D_{-s}^{\rm (up)*} &= D_{11,s} D_{11,-s}^* e^{2i\pi(B_2+B_3)} z_\infty^{2} , \\
D_s^{\rm (ref)} D_{-s}^{\rm (ref)*} &= D_{12,s} D_{12,-s}^* e^{2i\pi(B_2+B_3)} z_\infty^2 ,\\
D_s^{\rm (trans)} D_{-s}^{\rm (trans)*} &= e^{2i\pi(B_2+B_3)} (z_\infty-1)^2 .
\end{align}

In addition, from \eqref{CsDsconst1} and \eqref{CsDsconst2} we obtain the following relations:
\begin{align} 
\label{C22D11} \f{D_{11,s}}{C_{22,s}} &= \mk{ \f{z_r-1}{z_r} }^{2B_3} \mk{ \f{z_\infty-1}{z_\infty} }^{4s+2} 
\mk{\f{A}{A'}}^s F_s^{-1}, \\
\f{D_{12,s}}{C_{21,-s}^*} &= - \mk{ \f{z_\infty-1}{z_\infty} }^2 F_s^{-1}.
\end{align}
Here we implicitly assume that $C_{22,s}$ and $C_{21,-s}^*$ are nonvanishing.
If we consider $C_{22,s}=0$ for instance, then we should go back to \eqref{CDrel} and see $D_{11,s}=0$.

To summarize, we have solved the scattering problem exactly in the sense that we have expressed the coefficients for the ``in'' and ``up'' solutions in terms of the connection coefficients between the local Heun function, which is the exact solution of the Teukolsky equation.
Our calculation does not rely on any approximations such as the high-/low-frequency limit or slow-rotation limit.
A specific example is the WKB or eikonal approximation for the high-frequency regime, which is commonly used in the literature. 
Such approaches with approximations are helpful to extract a simple intuitive picture and formulae for a limited setup.
On the other hand, our exact formulation actually provides a simple expression without restriction of the parameter set.
This allows us to use a simple and fast computation to understand black hole physics, which we shall explore in \S\ref{sec:app}.

\section{Applications}
\label{sec:app}

In this section, we highlight several applications of the exact formulation of the scattering problem in \S\ref{sec:scat}.
We discuss the quasinormal modes in \S\ref{ssec:qnm}, the S-matrix and cross section in \S\ref{ssec:smat}, the reflection and absorption rates, greybody factor, and superradiant scattering in \S\ref{ssec:cc}, and the Green function in \S\ref{ssec:gfun}.
Our exact formulation serves as a simple and fast computational method with arbitrary high precision compared to the direct numerical integration of the Teukolsky equation.

\subsection{Quasinormal modes}
\label{ssec:qnm}

We can obtain QNM frequencies by requiring the regularity condition~\eqref{exactlambda} on the angular part, as well as the boundary condition on the radial part with a purely ingoing wave at the black hole horizon $r\to r_+$ and a purely outgoing wave at the cosmological horizon $r\to r'_+$.
Specifically, the condition on the radial part is given by $C_{22,s}=0$, i.e.,
\be \label{QNMfreq} W_z[y_{02,s}, y_{11,s}]=0. \ee
From \eqref{CDrel}, this condition is equivalent to $D_{11,s}=0$. 
It is also clear from \eqref{Rin-Heun} and \eqref{Rup-Heun} that 
$R_{{\rm in},s}$ with $C_{22,s}=0$ and $R_{{\rm up},s}$ with $D_{11,s}=0$ coincide with each other up to normalization.
For both cases one ends up with waves that satisfy the boundary condition for the QNM.
Note that, from the point of view of computational cost, the condition~\eqref{QNMfreq} is better than directly using $C_{22,s}=\f{W_z[y_{02,s}, y_{11,s}]}{W_z[y_{12,s}, y_{11,s}]}=0$ as the condition, since we do not need to calculate the Wronskian in the denominator.

In parallel to the angular condition~\eqref{exactlambda}, the radial condition~\eqref{QNMfreq} also depends on $\omega$ and $\lambda$ implicitly.
For a nonrotating black hole, we can plug in the eigenvalue $\lambda=\ell(\ell+1)-s(s-1)$ and solve \eqref{QNMfreq} only to obtain the QNM frequencies with a root-finding algorithm.
For a rotating black hole, we obtain $\omega$ and $\lambda$ by solving \eqref{exactlambda} and \eqref{QNMfreq} simultaneously.
The Wronskian is given by the exact solution in terms of the local Heun function, which in practice we can calculate by using the built-in function {\tt HeunG} implemented in Mathematica 12.1 or later.
This method gives us an arbitrary-precision arithmetic for the QNM frequencies.
As already shown in \cite{Hatsuda:2020sbn} for the Kerr-de Sitter black hole, this method is quite fast, typically within $\mO(1)$ second, 
and yields QNM frequencies that are consistent with the results in the literature.
Therefore, we do not repeat the calculation of the QNM frequencies here.
A caveat is that, to numerically find out the correct root, one needs to set an initial value sufficiently close to the root.

Let us note some technical details.
To optimize the calculation, one can choose the evaluation point of the Wronskians either within the overlapping region of both disks of convergence for the local Heun functions at $z=0$ and $1$, or outside but still near the overlapping region. 
While the radius of convergence for $y_{02,s}$ is always $1$ and does not depend on the parameters of the KNdS geometry or scattered waves, this is not the case for $y_{11,s}$.
Specifically, as one takes smaller $\Lambda M^2$, the radius of convergence for $y_{11,s}$ becomes smaller.
In such a case we find that $z=0.9$ is a convenient choice that yields a short computation time.

Let us note some differences from Leaver's method~\cite{Leaver:1985ax}.
Leaver's method is one of the most successful algorithms to calculate the QNM frequencies, and it is implemented in the Kerr-de Sitter black hole in \cite{Yoshida:2010zzb}.
Both Leaver's method and the above method yield QNM frequencies with arbitrary high precision without any approximations, and there are the following qualitative differences.
In Leaver's method, one solves three-term recurrence relations associated with the angular and radial equations in terms of infinite continued fractions.
The boundary conditions determine the eigenvalues and the QNM frequencies implicitly as roots of two equations containing infinite continued fractions.
One should then truncate the continued fractions appropriately, and use a root-finding algorithm.
In principle this procedure allows one to obtain an analytic expansion formula.
However, for high-precision computation one should take care with the convergence of the truncation.
On the other hand, for the root finding procedure for \eqref{exactlambda} and \eqref{QNMfreq}, the ambiguities of the truncation do not appear.
One can control the precision of the QNM frequencies solely by the precision of the calculation of the Wronskians.
However, the Wronskians are calculated numerically and their analytic expressions are unclear.
Therefore, the two arbitrary-precision arithmetics have complementary advantages.

\subsection{S-matrix and cross section}
\label{ssec:smat}

By definition \eqref{Rin-asym}, the ``in'' solution $R_{\text{in},s}$ is the solution satisfying the boundary condition 
with the purely ingoing boundary condition at the black hole horizon. 
We can then define the S-matrix~$\mS_{\ell, s}(\omega)$ as a ratio between the coefficients for the ingoing and outgoing waves at the cosmological horizon:
\be \label{Smat1} \mS_{\ell, s}(\omega) = (-1)^{\ell+1}\df{C_s^{\text{(ref)}}}{C_s^{\text{(inc)}}}. \ee
For numerical calculation, one can evaluate the S-matrix by numerically integrating the radial Teukolsky equation~\eqref{diffeq-Y} by requiring the boundary condition.
On the other hand, with the exact solution, we can use \eqref{eq:Cinc} and \eqref{eq:Cref} to obtain $C_s^{\text{(ref)}}/C_s^{\text{(inc)}}= A'^{2B_2+s}C_{21,s}/C_{22,s}$. 
Further, plugging $C_{21,s}, C_{22,s}$ into \eqref{C11}, we obtain 
\be \label{Smat2} \mS_{\ell, s}(\omega) 
= (-1)^{\ell+1} A'^{2B_2+s} \df{W_z[y_{12,s}, y_{02,s}]}{W_z[y_{02,s}, y_{11,s}]} .
\ee
Again, this expression reduces the number of Wronskians and minimizes the computational cost.
This formula does not require numerical integration.
We can obtain the S-matrix by calculating the ratio of the Wronskians of the local Heun function.

Given the S-matrix, we can write down the differential cross sections and the scattering amplitudes, which are defined through an infinite series of the partial wave expansion.
In practice, one needs to truncate the infinite series at some finite $\ell_{\rm max}$.
However, it is known that a naive truncation of the partial wave expansion introduces a numerical error, and hence special care is required~\cite{Dolan:2008kf}.
The situation is analogous to the computation of the Coulomb scattering series~\cite{PhysRev.95.500}.
Since our main goal in the present paper is to establish the analytic formulation of the wave scattering from black holes, here we do not address this issue further.

\subsection{Reflection and absorption rates}
\label{ssec:cc}

We can express the conserved current of the scattered wave~\cite{Teukolsky:1974yv} in terms of the exact solution.
We shall see below that the exact formulation provides a simple formula for the reflection rate and absorption (transmission) rate or the greybody factor.

As we explained above, $R_s$ and $\Delta^{-s}R_{-s}^*$ are linearly independent solutions of the same differential equation~\eqref{radTeu}.
Plugging $(R_1,R_2)=(R_{{\rm in},s},\Delta^{-s}R_{{\rm in},-s}^*)$ and $(R_{{\rm up},s},\Delta^{-s}R_{{\rm up},-s}^*)$ into \eqref{RdRconst} and evaluating it at $r\to r_+$ and $r\to r'_+$, we obtain 
\begin{align} 
C_s^{\rm (inc)}C_{-s}^{\rm (inc)*} &= C_s^{\rm (ref)}C_{-s}^{\rm (ref)*} + F_s C_s^{\rm (trans)}C_{-s}^{\rm (trans)*} , \\
D_s^{\rm (up)}D_{-s}^{\rm (up)*} &= D_s^{\rm (ref)}D_{-s}^{\rm (ref)*} + F_s^{-1} D_s^{\rm (trans)}D_{-s}^{\rm (trans)*} .
\end{align}
These relations imply energy conservation~\cite{Teukolsky:1974yv}.
While we do not specify the relative normalization between the $s$ and $-s$ solutions, the normalization degrees of freedom do not enter if we write down the energy conservation in the form
\begin{align}
\label{cceq1} 1 &= \f{C_s^{\rm (ref)}C_{-s}^{\rm (ref)*}}{C_s^{\rm (inc)}C_{-s}^{\rm (inc)*}} + F_s \f{C_s^{\rm (trans)}C_{-s}^{\rm (trans)*}}{C_s^{\rm (inc)}C_{-s}^{\rm (inc)*}} , \\
\label{cceq2} 1 &= \f{D_s^{\rm (ref)}D_{-s}^{\rm (ref)*}}{D_s^{\rm (up)}D_{-s}^{\rm (up)*}} + F_s^{-1} \f{D_s^{\rm (trans)}D_{-s}^{\rm (trans)*}}{D_s^{\rm (up)}D_{-s}^{\rm (up)*}} .
\end{align}
The physical meaning is transparent. 
The first term on the right-hand side of \eqref{cceq1} indicates the probability of the incoming wave being reflected by the black hole, whereas the second term means the probability of the incoming wave transmitting the effective potential and falling into the black hole.
Therefore, the first and second terms yield the reflection rate $\mR_s$ and transmission rate $\mT_s$, respectively.
Note that, in the context of black hole scattering, $\mT_s$ is also called the absorption rate since transmission through the effective potential means absorption by the black hole.
Similar logic also holds for each term on the right-hand side of \eqref{cceq2}.
In particular, the second term on the right-hand side of \eqref{cceq2} is the greybody factor $\Gamma_s$, which is the probability of the outgoing wave reaching the cosmological horizon.
With \eqref{CsDsconst1} and \eqref{CsDsconst2}, we can see that \eqref{cceq1} and \eqref{cceq2} are equivalent.
Namely, we can rewrite them as 
\be \label{RT1} \mR_s + \mT_s = 1, \ee
where  
\begin{align}
\label{cdcc1} \mR_s &\equiv \f{C_s^{\rm (ref)}C_{-s}^{\rm (ref)*}}{C_s^{\rm (inc)}C_{-s}^{\rm (inc)*}} = \f{D_s^{\rm (ref)}D_{-s}^{\rm (ref)*}}{D_s^{\rm (up)}D_{-s}^{\rm (up)*}} , \\
\label{cdcc2} \mT_s &\equiv F_s \f{C_s^{\rm (trans)}C_{-s}^{\rm (trans)*}}{C_s^{\rm (inc)}C_{-s}^{\rm (inc)*}} = F_s^{-1} \f{D_s^{\rm (trans)}D_{-s}^{\rm (trans)*}}{D_s^{\rm (up)}D_{-s}^{\rm (up)*}} \equiv \Gamma_s.
\end{align}
The relation~\eqref{cdcc2} guarantees that the absorption rate~$\mT_s$ coincides with the greybody factor~$\Gamma_s$.

Using the exact solution, we can derive the following simple expression:
\be \label{refrate} \mR_s = \df{W_z[y_{12,s}, y_{02,s}]}{W_z[y_{02,s}, y_{11,s}]} \mk{ \df{W_z[y_{12,-s}, y_{02,-s}]}{W_z[y_{02,-s}, y_{11,-s}]} }^* , \ee
where we have used \eqref{Smat1} and \eqref{Smat2}.
The absorption rate is then given by $\mT_s=1-\mR_s$. 
Alternatively, from \eqref{Cincsq} and \eqref{Ctranssq}, we obtain 
\be \label{trnrate} \mT_s = F_s \mk{\f{z_\infty}{z_\infty-1}}^2 \f{1}{C_{22,s}C_{22,-s}^*} . \ee
By virtue of \eqref{cdcc2}, these formulae also allow us to calculate the greybody factor $\Gamma_s$.

In particular, for the scalar wave $s=0$, the absorption rate can be written as
\be \mT_0 = F_0 \mk{\f{z_\infty}{z_\infty-1}}^2 \f{1}{|C_{22,0}|^2} . \ee
Therefore, the absorption rate can be negative if $F_0=J(r_+)/J(r'_+)<0$ is satisfied.
This is nothing but superradiant scattering. 
The condition $J(r_+)/J(r'_+)<0$ can be rewritten as
\be \label{superrad} \f{am+\f{eQr'_+}{1+\alpha}}{r'^2_++a^2} < \omega < \f{am+\f{eQr_+}{1+\alpha}}{r^2_++a^2} . \ee

Here we stress that we have obtained these formulae exactly without any approximations, e.g., the high-/low-frequency limit.
Recalling that $z_\infty$ and $F_s$ defined in \eqref{Fs} can be algebraically obtained, 
the only necessary calculation that one needs to perform to obtain \eqref{refrate} or \eqref{trnrate} is the evaluation of the Wronskians between two local Heun functions at $z=0$ and $1$, which is achievable within the overlapping region of the two disks of convergence.
Furthermore, compared to the calculation for the QNM frequencies in \S\ref{ssec:qnm}, the calculation for the reflection/absorption rate does not require an initial value close to the solution.

\begin{figure}[H]
  \centering
  \includegraphics[width=0.45\linewidth]{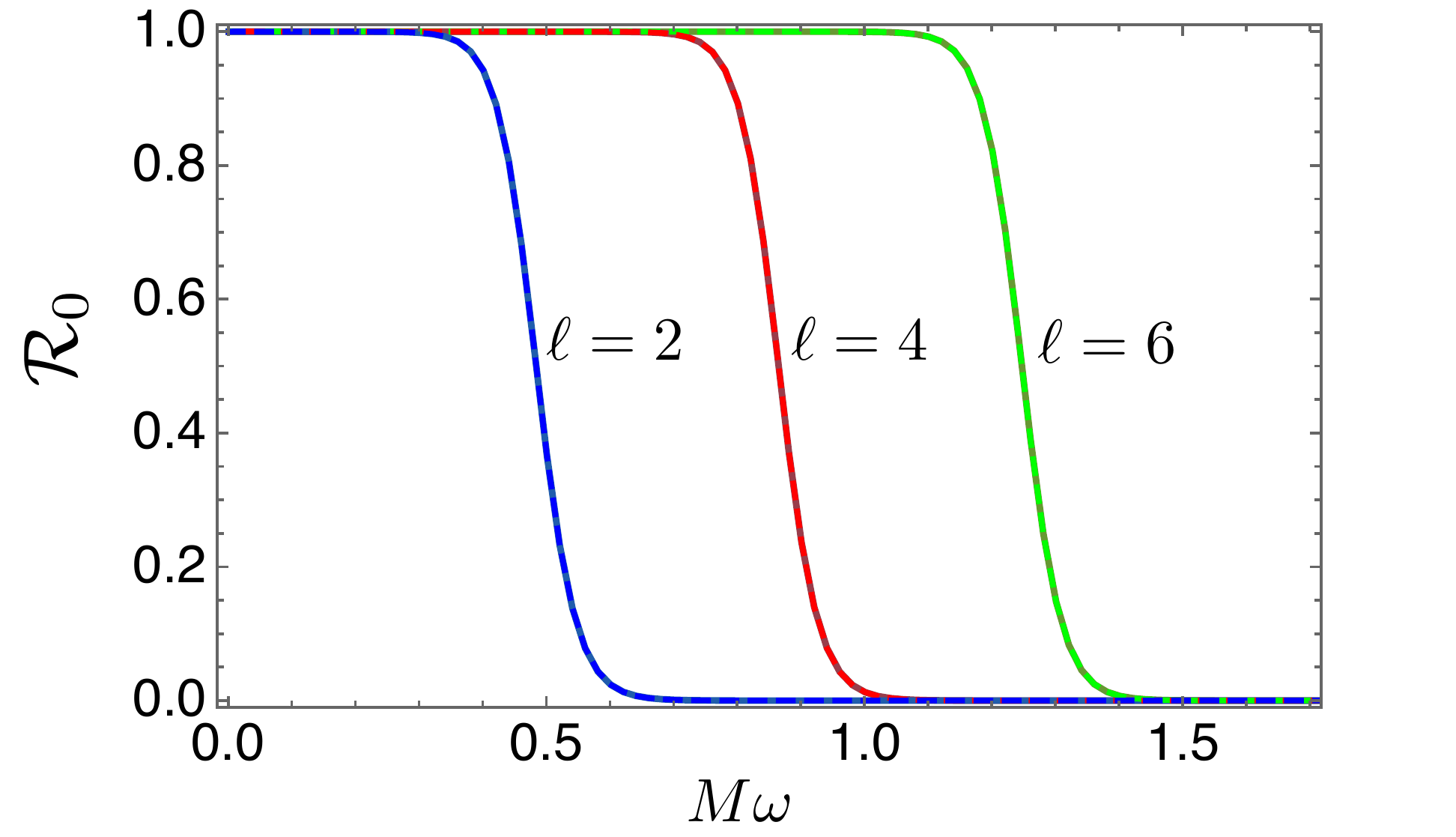}
  \caption{\footnotesize{The reflection rate $\mR_s$ for the scalar wave $(s=0)$ with $\ell=2$~(blue), $4$~(red), $6$~(green) by the SdS black hole with $\Lambda M^2=10^{-3}$ obtained by the exact formula~\eqref{refrate} (solid) and numerical integration (dashed).}}
  \label{fig:R_scalar}
\end{figure}

\begin{figure}[H]
  \centering
  \includegraphics[width=0.96\linewidth]{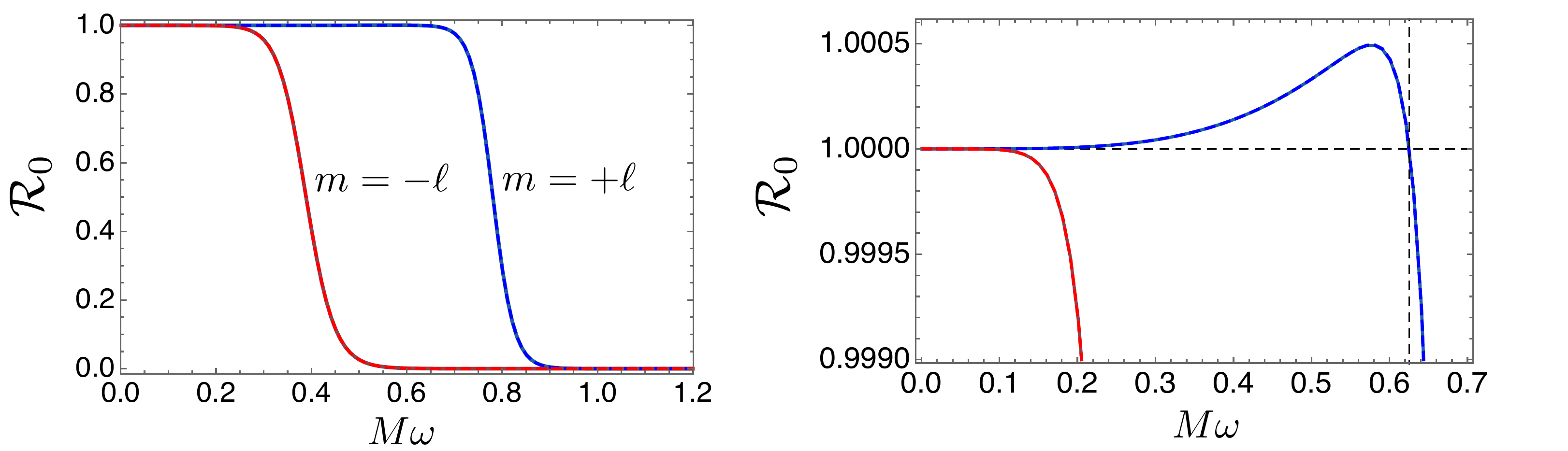}
  \caption{\footnotesize{Left: The reflection rate $\mR_s$ for the scalar wave $(s=0)$ of $m=+\ell$~(blue) and $m=-\ell$~(red) modes with $\ell=2$ scattered by the Kerr-de Sitter black hole with $a/M=0.9$ and $\Lambda M^2=10^{-3}$ obtained by the exact formula~\eqref{refrate} (solid) and numerical integration (dashed).  
  Right: Close-up plot of the left panel to highlight the superradiance.  The dashed vertical line is the upper bound of the superradiant frequency~\eqref{superrad}. }}
  \label{fig:R_scalar_Kerr}
\end{figure}

In Fig.~\ref{fig:R_scalar}, we present the reflection rate $\mR_s$ of the scalar wave with $\ell=2,4,6$ by the SdS black hole with $\Lambda M^2=10^{-3}$ as a function of $M\omega$. 
We calculated $\mR_s$ for $0\leq M\omega \leq 1.7$ with the sampling mesh size $\Delta(M\omega)=0.02$.
The solid curves are the results obtained by the exact formula~\eqref{refrate}.
As a consistency check, we also calculated the reflection rate numerically using Mathematica.
First, using {\tt NDSolve} with the method ``{\tt StiffnessSwitching}'', we numerically integrate the radial Teukolsky equation in the Schr\"odinger form~\eqref{diffeq-Y} with the ingoing boundary condition near the BH horizon for ${\cal{Y}}_{s}$, which is the negative sign of \eqref{Yasym}.
We then fit the behavior of the obtained wave function for the large-$r_*$ region, where the effective potential converges as \eqref{Vasym}, by the asymptotic solutions and read off the coefficients for the in/outgoing waves. 
Here we use the asymptotic solutions~\eqref{Yasym} in terms of the tortoise coordinate $r_*$ rather than $r$ since the frequency of oscillation diverges in $r$ space. 
Note that the numerical calculation is done with the default machine precision. 
For the sake of clarity, let us note the specifications of our computer for the numerical calculation. 
We use a Mac Pro with a 3~GHz, 8-core processor and the command {\tt ParallelTable} is used to get the list of data. 
For all the numerical calculations for the reflection rate in the present paper (Figs.~\ref{fig:R_scalar} and \ref{fig:R_scalar_Kerr}), we adopt the above method.

The results of the numerical integration are shown by dashed curves in Fig.~\ref{fig:R_scalar} and are in good agreement with the results of the exact formula shown by solid curves. 
While analytic calculations known in the literature are valid under certain approximations such as high/low multipoles, our exact formula is based on the exact solution without approximation and hence can be used for a wider range of multipoles $\ell$.
The result in Fig.~\ref{fig:R_scalar} is also consistent with physical intuition since partial waves with $\ell\lesssim \ell_c$ are absorbed by black holes, where $\ell_c=3\sqrt{3}M\omega$ is the critical angular momentum.
This is because partial waves with the impact parameter $\sim b_c\equiv \ell_c/\omega= 3\sqrt{3}M$ are marginally scattered at the vicinity of the peak of the effective potential.

There are several differences between the exact formula~\eqref{refrate} and the numerical calculation.
First of all, the exact formula allows us to obtain the reflection rate with arbitrary high precision.
One can easily improve the precision by requiring higher precision for the root-finding algorithm.  
On the other hand, for the numerical integration, it is more difficult to control the precision since one needs to take account of several processes to solve the differential equation. 
For the above calculations, the computation time for the exact formula is comparable to the numerical calculation.
Specifically, to obtain each curve in Fig.~\ref{fig:R_scalar}, it takes about 10~sec for the numerical calculation without setting {\tt PrecisionGoal} in Mathematica (the machine precision) and about 20~sec for the exact formula with 
{\tt PrecisionGoal}~$\rightarrow 15$, respectively.

For scalar waves $(s=0)$ scattered by the Kerr-de Sitter black hole ($Q=0$, $a/M=0.9$, $\Lambda M^2=10^{-3}$), we plot the reflection rate $\mR_s$ of $m=\pm \ell$ modes with $\ell=2$ in Fig.~\ref{fig:R_scalar_Kerr}. 
For the eigenvalue $\lambda$, instead of solving the angular part, we use the analytic expansion formula~\cite{Suzuki:1998vy} since the error remains $\mO(10^{-1})$\% for this setup, as shown in Fig.~\ref{fig:lambdas}.
We obtain the solid curves by using the exact formula~\eqref{refrate}, and the dashed curves by numerical integration.
The sampling mesh size is $\Delta(M\omega) =0.01$; i.e., we take $120$ points in the range $0\leq M\omega\leq 1.2$.
We used {\tt PrecisionGoal}~$\to 15$ for the exact formula, and the default machine precision for the numerical calculation.
Using {\tt Table} to list the data, the computation time to get each curve in Fig.~\ref{fig:R_scalar_Kerr} is 100~sec for numerical calculation and 240~sec for the exact formula. 
We see that the two results are in good agreement.
The blue (red) curve depicts $m=\ell$ ($m=-\ell$), corresponding to the case where the angular momentum of the black hole and incident wave are (oppositely) aligned. 
The difference in the alignment causes the difference in the critical impact parameter $b_c \equiv \ell_c/\omega$ at which a transition occurs from absorption to reflection.

In the right panel of Fig.~\ref{fig:R_scalar_Kerr}, we show a closer look of the left panel to confirm the superradiant scattering for $m=+\ell$. 
Indeed, we can see that the reflection rate exceeds unity, shown by the horizontal dashed line.  
For this parameter set, the condition~\eqref{superrad} on the superradiant frequency reads $6.23\times 10^{-4}<M\omega<6.25\times 10^{-1}$.
The vertical dashed line corresponds to the upper bound of the superradiant freqeuncy, which is consistent with our calculations.

\subsection{Green function}
\label{ssec:gfun}
\begin{figure}[H]
\centering
\includegraphics[width=0.65\linewidth]{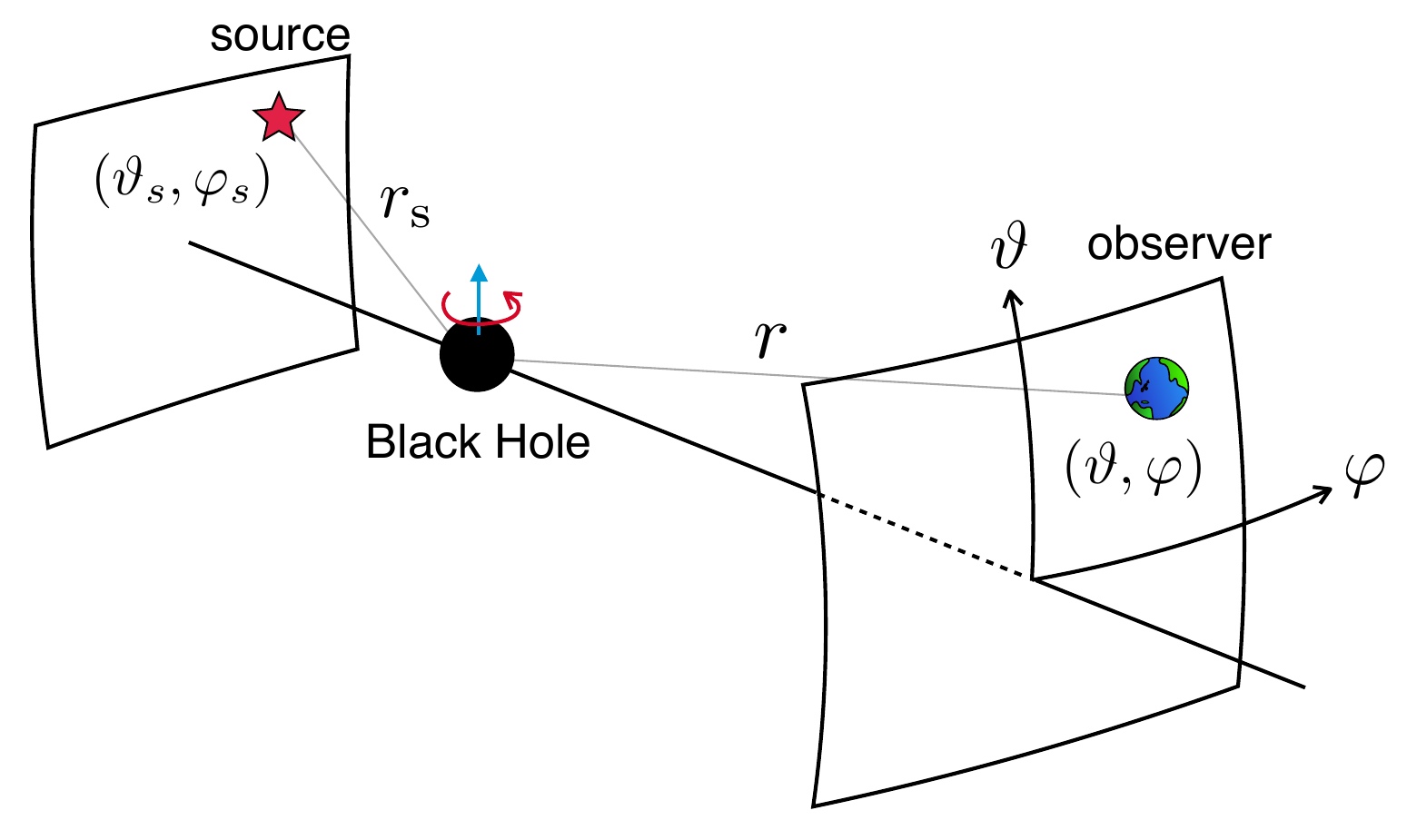}
\caption{\footnotesize{The configuration of the wave scattering problem.}}
\label{fig:conf}
\end{figure}

In this section, we construct the Green function for the wave scattering problem by a KNdS black hole in terms of the local Heun function. 
We choose the KNdS black hole as the origin of the spherical coordinate system $(r,\theta,\varphi)$ with the rotation axis at $\theta=0$. 
We assume a stationary point source, whose spatial location is denoted by $\bm{x}_{\rm s}=(r_{\rm s},\vartheta_{\rm s},\vp_s)$, and the observing point at $\bm{x}=(r,\vartheta,\vp)$, 
where $\vartheta$ is related to the polar angular variable $\theta$ of the spherical coordinates as $\vartheta=\pi/2-\theta$.
Therefore, $\vartheta=0$ is the equatorial plane of the KNdS black hole.
The relationship between these points and the black hole is shown in Fig.~\ref{fig:conf}. 

For the case where a spin-$s$ wave is emitted by a stationary point source, the spatial part of the Green function $G(\bm{x},\bm{x}_{\rm s})$ can be expanded with the partial waves as
\be
G(\bm{x},\bm{x}_{\rm s}) = \sum_{\ell=0}^\infty \sum_{m=-\ell}^{\ell} \tilde{G}_{\ell}(r,r_{\rm s}) _{s}S_{\ell m}(\theta) _{s}S_{\ell m}^{*}(\theta_{\rm s}) e^{im\vp}e^{-im\vp_\text{s}},
\label{eq:Green}
\ee
where $_{s}S_{\ell m}(\theta, \vp)$ is the modified spin-weighted spheroidal harmonics due to the presence of the cosmological constant, which can be expressed by the local Heun function as given in Eq.~\eqref{Sdef}.

The differential equation that the radial part $\tilde{G}(r, r_{\rm s})$ obeys can be derived from the master equation for the spin-$s$ wave with Dirac's $\delta$ function as the source term as
\be \label{eq:inhomo}
\Biggl[ \Delta^{-s}\f{d}{dr}\Delta^{s+1}\f{d}{dr} 
+ \f{J^2-isJ\Delta'}{\Delta} +2 i s J' 
- \f{2\alpha}{a^2}(s+1)(2s+1)r^2
+2s(1-\alpha)-\lambda \Biggr] \tilde{G}_{\ell}(r,r_{\rm s}) = -\delta (r-r_\text{s}) .
\ee
As we discussed in \S\ref{sec:scat}, the homogeneous equation~\eqref{radTeu} can be exactly solved in terms of the local Heun function, and satisfies the relation~\eqref{RdRconst}.
Following the standard prescription, we can construct the Green function by using the two linearly independent solutions and the constant~\eqref{RdRconst}, which is given by 
\be
 \tilde{G}_{\ell}(r,r_{\rm s})=
\dfrac{-\Delta^s(r_{\rm s})}{\Delta^{s+1} W_r[R_\text{in},R_\text{up}]}\left\{R_\text{in}(r_{\rm s})R_\text{up}(r)\Theta(r-r_{\rm s})+R_\text{in}(r) R_\text{up}(r_{\rm s})\Theta(r_{\rm s}-r)\right\},
\label{eq:rad_G}
\ee
where $\Theta(r)$ is the unit step function. 
Here, we have chosen $R_\text{in}$ and $R_\text{up}$ given in \eqref{Rin-Heun} and \eqref{Rup-Heun}, respectively, as a suitable pair of independent solutions to the radial Teukolsky equation by considering the boundary condition of the wave scattering problem by black holes. 
Note that we omit the subscript $s$ of $R_{{\rm in}, s}$ and $R_{{\rm up}, s}$ for the spin-$s$ wave for simplicity.
The denominator $\Delta^{s+1} W_r[R_\text{in},R_\text{up}]$ is the constant given in \eqref{RdRconst}, which should be evaluated at some $r$ between $r_+<r<r'_+$, or $0<z<1$.
Plugging \eqref{eq:rad_G} into \eqref{eq:Green}, we obtain the spatial part of the Green function as
\be
G(\bm{x},\bm{x}_{\rm s})=\sum_{\ell=0}^\infty \sum_{m=-\ell}^{\ell}\dfrac{-\Delta^s(r_{\rm s})\left\{R_\text{in}(r_{\rm s})R_\text{up}(r)\Theta(r-r_{\rm s})+R_\text{in}(r) R_\text{up}(r_{\rm s})\Theta(r_{\rm s}-r)\right\}}{\Delta^{s+1} W_r[R_\text{in},R_\text{up}]} \ _{s}S_{\ell m}(\theta) _{s}S_{\ell m}^{*}(\theta_{\rm s}) e^{im\vp}e^{-im\vp_\text{s}}.
\label{eq:Green_full}
\ee
In particular, as mentioned in \S\ref{ssec:ang}, for the scattering of scalar waves by a nonrotating black hole, the angular solution is given by the spherical harmonics.
In this case we can use the addition theorem for the spherical harmonics:
\be 
\sum_{m=-\ell}^{m=\ell}Y_{\ell m}(\theta,\vp)Y_{\ell m}^* (\theta_\text{s},\vp_\text{s})=\frac{2\ell+1}{4\pi} P_{\ell}(\cos{\gamma}), 
\ee 
where the variable $\gamma$ represents the angle between source and observer, which is defined by $\cos{\gamma}=\cos{\theta}\cos{\theta_{\rm s}}+\sin{\theta}\sin{\theta_{\rm s}}\cos{(\vp-\vp_{\rm s})}$.
We then arrive at
\be
G(\bm{x},\bm{x}_{\rm s})=\sum_{\ell=0}^\infty \f{2\ell+1}{4\pi}\dfrac{-\Delta^s(r_{\rm s})\left\{R_\text{in}(r_{\rm s})R_\text{up}(r)\Theta(r-r_{\rm s})+R_\text{in}(r) R_\text{up}(r_{\rm s})\Theta(r_{\rm s}-r)\right\}}{\Delta^{s+1} W_r[R_\text{in},R_\text{up}]} \ P_\ell\left(\cos{\gamma}\right).
\label{eq:Green_full2}
\ee

Let us check whether this formula reproduces the formula derived with asymptotic solutions of scalar fields ($s=0$) in Schwarzschild spacetime \cite{Nambu2016}. 
We use the rescaled radial function $\mY_s$~\eqref{def-Y} with $a=0$ and $s=0$, and the tortoise coordinate $r_*$. 
The relationship between the Wronskian for $R_s$ with respect to $r$ and that for $\mY_s$ with respect to $r_*$ is then given by
\be
\Delta W_{r}[R_{\text{in}},R_{\text{up}}] = W_{r_*}[{\cal{Y}}_{\text{in}},{\cal{Y}}_{\text{up}}].
\ee
Plugging this into the Green function \eqref{eq:Green_full2} yields 
\be
G(\bm{x},\bm{x}_{\rm s})= -\sum_{\ell=0}^{\infty}\df{2\ell+1}{4\pi r r_{\rm s}}
\df{{\cal{Y}}_{\text{in}}(r_{\rm s}) {\cal{Y}}_{\text{up}}(r)\Theta(r-r_{\rm s})+{\cal{Y}}_{\text{in}}(r) {\cal{Y}}_{\text{up}}(r_{\rm s})\Theta(r_{\rm s}-r)}{ W_{r_*}[{{\cal{Y}}_{\text{in}},{\cal{Y}}_{\text{up}}]}} \ P_\ell\left(\cos{\gamma}\right),
\label{eq:Green_Sch}
\ee 
which amounts to the Green function derived in \cite{Nambu2016}.

Let us highlight several differences between the exact Green function~\eqref{eq:Green_full} and the analysis performed in \cite{Nambu2016}.
First, in \cite{Nambu2016} the Green function \eqref{eq:Green_Sch} was evaluated by substituting the asymptotic forms of the radial function corresponding to \eqref{Rin-asym} and \eqref{Rup-asym}. 
However, in that case the sum over the partial waves does not converge due to $1/r$ behavior of the gravitational potential. 
This issue originates from the use of the asymptotic solutions.  
Indeed, in \cite{Nambu2016}, the convergence issue was circumvented by adding a finite-distance correction to the asymptotic solutions, which play the role of regulator for the partial wave sum.
In contrast, for the exact Green function~\eqref{eq:Green_full}, there is no convergence issue intrinsically.
This is because the exact Green function does not rely on the asymptotic solution or approximations, and inherits finite-distance effects at nonlinear order.

Second, in \cite{Nambu2016}, it is assumed that both the source and observer are located at a sufficiently distant $r/M,r_{\rm s}/M\gg 1$, but are finite points so that one can substitute the asymptotic solution with the correction term into the Green function. 
Furthermore, the wavelength of the scalar wave is restricted to the short-wavelength case $M\omega\gg 1$ to evaluate the phase shift within the WKB approximation.
A small deflection angle ($\vartheta \sim 0$, $\vp \sim 0$) was additionally assumed to obtain a simple formula.
In contrast, in our formulation there is no approximation and no restriction on the scattered wave and the configuration of the source and the observer since the radial functions $R_{{\rm in/up}}$ represented in terms of the local Heun function are the exact solution to the radial Teukolsky equation.
Moreover, our Green function~\eqref{eq:Green} applies to a more general case, i.e., spin $0, \f{1}{2}, 1, \f{3}{2}, 2$ massless fields on the Kerr-de Sitter background and those for spin $0,\f{1}{2}$ massless fields on the KNdS background.
Therefore, the Green function \eqref{eq:Green_full} is the most general exact formula for wave scattering by a KNdS black hole.

\begin{figure}[H]
  \centering
 \includegraphics[width=0.96\linewidth]{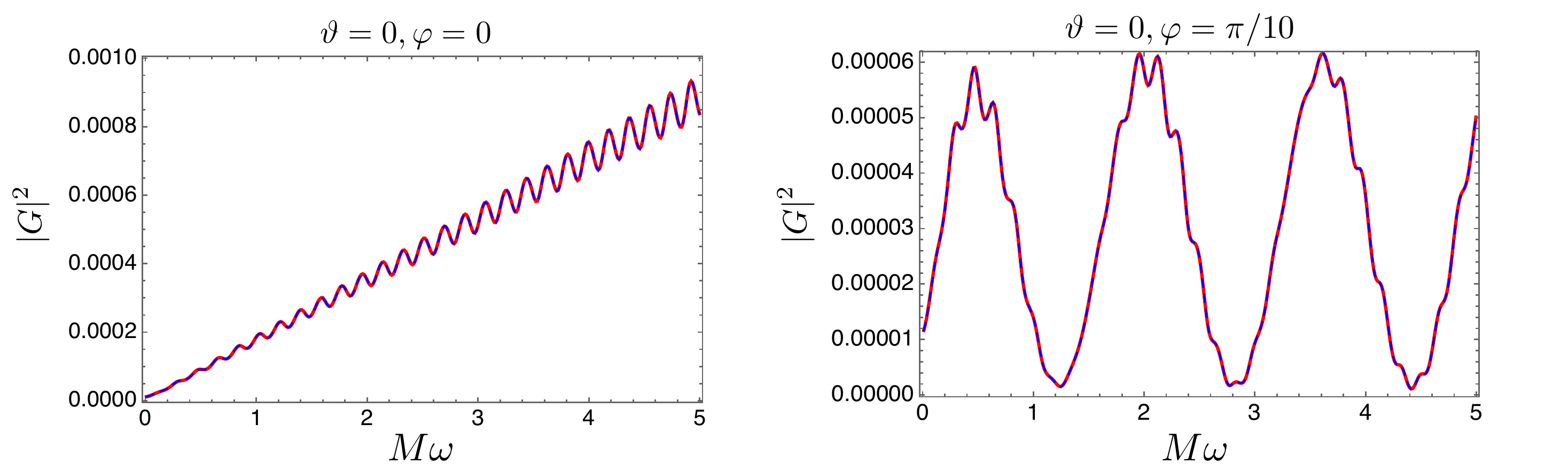}
 \caption{\footnotesize{The power spectrum of the scalar wave $s=0$ emitted from the source at $(r_{\rm s}, \vartheta_{\rm s}, \vp_{\rm s})=(6M,0,\pi)$ and scattered by the SdS black hole with $\Lambda M^2=10^{-3}$.
 The observer is located at the exactly forward direction $(r,\vartheta,\vp)=(20M,0,0)$ (left) and slightly off forward direction $(r,\vartheta,\vp)=(20M,0,\pi/10)$ (right).
 The Green function~\eqref{eq:Green_full2} is obtained by the exact solution in terms of the local Heun function (solid red) and the numerical integration of the radial Teukolsky equation (dashed blue), respectively.}}
 \label{fig:power}
\end{figure}

In Fig.~\ref{fig:power}, we present the power spectrum, i.e., the absolute square of the Green function~\eqref{eq:Green_full2} measured at $r=20M$ for forward scattering by the SdS black hole with $\Lambda M^2=10^{-3}$ of scalar waves emitted from the source located at $(r_{\rm s}, \vartheta_{\rm s}, \vp_{\rm s})=(6M,0,\pi)$.
To obtain the power spectrum, we evaluate $R_{{\rm in/up}}$ in the following two ways: 
First, we employ the exact solution \eqref{Rin-Heun}, \eqref{Rup-Heun} in terms of the local Heun function with {\tt PrecisionGoal}~$\rightarrow 25$, which is shown as solid red curves in Fig.~\ref{fig:power}.
On the other hand, the power spectrum obtained by the numerical integration is shown by dashed blue curves.  
For the numerical integration, here we improve a similar calculation performed in \cite{Nambu2019}.
In \cite{Nambu2019}, the WKB approximation was partially employed, but here we do not use the approximation. 
Here, to obtain $R_{{\rm in/up}}$, we numerically integrate the differential equations~\eqref{diffeq-Y} and \eqref{drstardef} with the boundary conditions~\eqref{Rin-asym} and \eqref{Rup-asym}.
We choose the location to impose the boundary conditions sufficiently close to each horizon, and confirm that the results are almost unaffected by some change of the location.
Specifically, we start the numerical integration with the purely ingoing boundary condition as the initial condition at a nearby point of the black hole horizon $r_\text{i}/M=r_{+}/M+10^{-6}\sim 2.0027$ and solve the radial equation towards the source point $r_\text{s}/M=6$, 
whereas, for $R_\text{up}$, the radial equation is solved with a purely outgoing boundary condition from a point near the de Sitter horizon $r_\text{f}/M= r'_{+}/M-0.74\sim 53$ to $r_\text{s}/M=6$. 
Then, substituting these solutions into~\eqref{eq:Green_full2}, the Green function is obtained.
To get both results, we truncate the partial wave sum at $\ell_\text{max}=8M\omega+6$, which we find yields a good convergence.
As shown in Fig.~\ref{fig:power}, the two curves are in good agreement. 
The computational time to obtain the curves in Fig.~\ref{fig:power} is about 30 min for the exact formula and 1 min for the numerical integration.
While the exact formula takes longer, note that the numerical integration here does not have high precision.
Actually, the numerical result matches the exact result up to 3 digits only.
To improve the numerical result, one needs to choose the location for the boundary condition closer to each horizon, and to require higher precision for the root-finding algorithm and the differential equation solver.
For instance, if we take $r_\text{f}/M= r_{+}^{\prime}/M-10^{-6}$, the numerical result matches the exact result up to 5 digits, and the computational time is 3 min in this case.
It would thus be fair to say that the exact formula serves a simple calculation method with high precision.
It allows an arbitrary high-precision calculation and it is easier to control the precision without numerically solving the differential equation.

For exact forward scattering with $(r,\vartheta,\vp)=(20M,0,0)$ in the left panel of Fig.~\ref{fig:power}, the behavior that $|G|^2$ increases linearly stems from the property of the caustics at the forward position of the present scattering problem, which will diverge for $M\omega \rightarrow \infty$.
The period of oscillation on the linear growth reflects the scale of the peak of the effective potential.
This corresponds to the position of the unstable circular photon orbit in the geometrical optics limit, and is evaluated as $M \Delta \omega \sim 1/(3\sqrt{3}) \sim 0.2$ for $\Lambda M^2 \ll 1$. 
On the other hand, for the case of the slightly off forward scattering with $(r,\vartheta,\vp)=(20M,0,\pi/10)$ in the right panel of Fig.~\ref{fig:power}, there is one more oscillating scale with a longer period. 
This originates from the breaking of the symmetry of the relative relation of the source-black hole-observer positions, 
which causes interference due to the difference of light ray paths in the limit of the geometrical optics. 

\begin{figure}[H]
  \centering
  \includegraphics[width=0.65\linewidth]{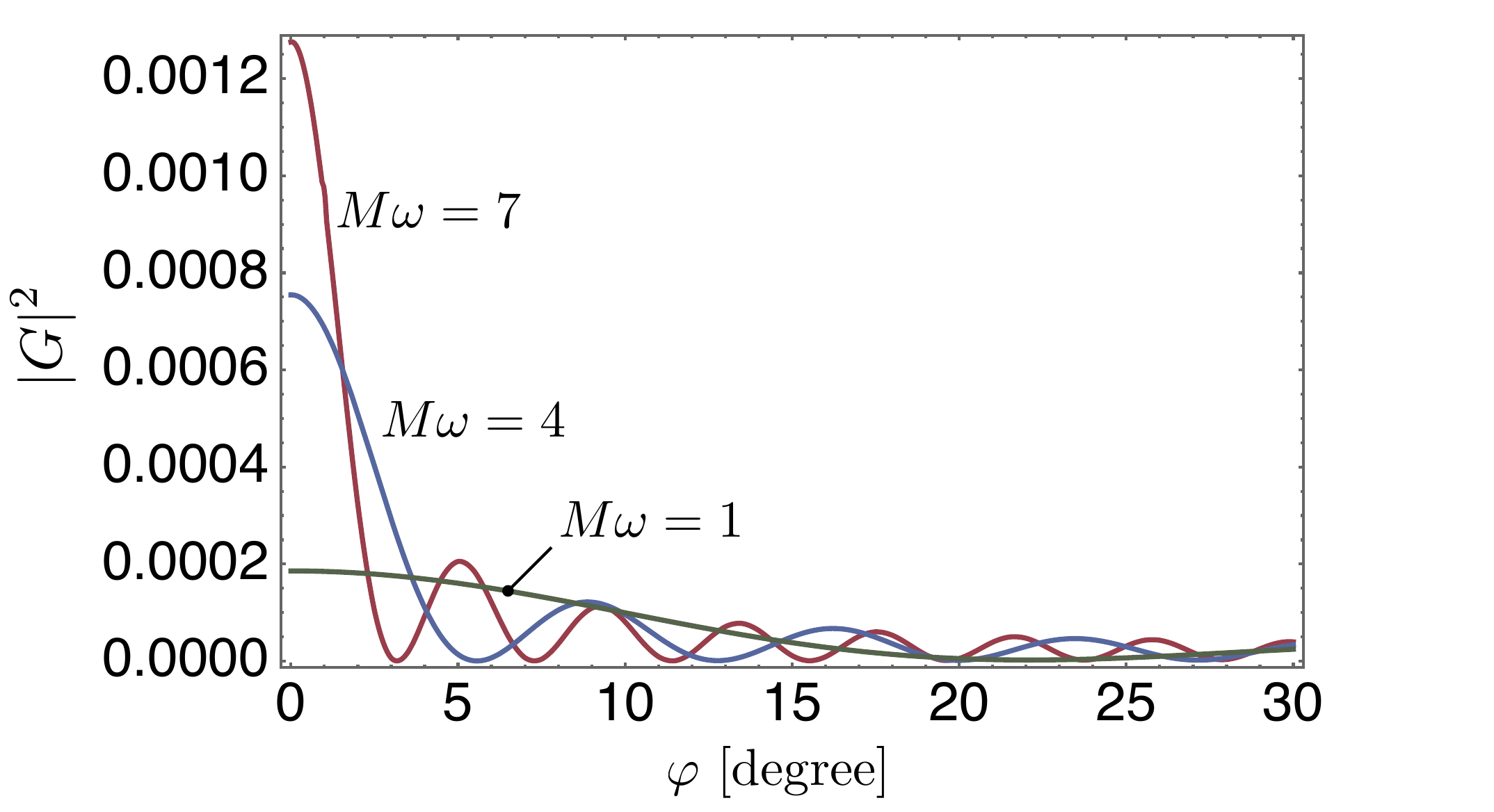}
  \caption{\footnotesize{Angular dependence of the absolute square of the exact Green function~\eqref{eq:Green_full2} for the scattered scalar waves for fixed frequency $M\omega=7$, $4$, $1$ with $(r,\vartheta)=(20M,0)$ and $(r_{\rm s},\vartheta_{\rm s},\vp_{\rm s})=(6M,0,\pi)$.}}
  \label{fig:int1d}
\end{figure}

As another demonstration, we present the angular dependence of the absolute square of the scattered scalar wave for fixed frequency $M\omega=7$, $4$, $1$ in Fig.~\ref{fig:int1d}. 
As expected, it shows a peak at $\vp=0$ and decays with oscillations depending on the fixed frequency.
We see that our exact formula is valid for a wide range of the azimuthal angle.

We have provided several examples of scalar wave scattering by the SdS black hole with a small cosmological constant and compared the results with previous works.
Since the main goal of the present paper is to establish the formulation, we have avoided to present too many specific calculations.
However, our formula~\eqref{eq:Green_full} is quite general and applies to the wave scattering of the spin-$s$ field from the KNdS black hole. 
We will investigate the details of several observables in wave optical gravitational lensing for a more general case in a future work.

\section{Conclusion}
\label{sec:con}

In this paper we have established the exact formulation for the wave scattering problem by the KNdS black hole.
We consider the propagation of a massless field with spin and charge on the KNdS background.
The Teukolsky equations for spin $0, \f{1}{2}, 1, \f{3}{2}, 2$ fields on the Kerr-de Sitter background and those for spin $0,\f{1}{2}$ fields on the KNdS are separable and take the unified form.
Here, the spin 0 field corresponds to a scalar field conformally coupled to gravity.
Transforming the angular and radial Teukolsky equations into Heun equations, we can write down the exact solution in terms of the local Heun functions at regular singular points. 
For the angular solution, we can impose the regularity condition by requiring the linear dependence of the local Heun functions. 
For the radial equation, with the appropriate transformation, we can respectively map the black hole horizon $r=r_+$ and the cosmological horizon $r=r'_+$ into $z=0$ and $1$, and discuss the scattering problem within the range $0\leq z\leq 1$.
For this setup, there exists an overlapping region of the two disks of convergence of the local Heun functions at $z=0$ and $1$, and we can discuss the scattering problem.
We can write down the ``in'' and ``up'' solutions, which satisfy certain boundary conditions, in a fully analytic way without any approximations.
We have expressed the coefficients for the asymptotic in/outgoing waves exactly in terms of the connection coefficients for the local Heun functions at $z=0$ and $1$, which are given as the ratio of the Wronskians of the local Heun function.
Once the coefficients are obtained exactly, we can write down various important quantities for black hole scattering exactly.

We have highlighted several applications of our exact formulation.
It has already been shown in \cite{Hatsuda:2020sbn} that the local Heun function is a powerful tool to calculate the QNM frequencies for the Kerr-de Sitter black hole with arbitrary high precision.
Given a sufficiently close initial value as an input, one can obtain the QNM frequencies very quickly.
We have generalized this result to the KNdS geometry and provided the arbitrary-precision arithmetic for the KNdS QNM frequencies for spin $0,\f{1}{2}$ massless fields for the first time.
We can also write down the S-matrix, with which the differential cross sections and the scattering amplitudes can be written down.
Further, we have explored the conserved current for the scattering problem in terms of the exact solution, and derived simple formulae for the reflection/absorption rate and the greybody factor (see also \cite{Gregory2020} for a recent study on the greybody factor and Hawking radiation for the Kerr-de Sitter black hole in this context).
We have checked the consistency between the results obtained by our exact formula and numerical integration, and clarified the efficiency of our formula for the reflection rate in comparison with the numerical integration. 
Finally, we have constructed the Green function for the wave scattering from the KNdS black hole.
We have calculated the power spectrum as the absolute square of the Green function to see the frequency dependence of the forward and slightly off forward scattering, as well as the angular dependence for the fixed frequency waves. 
They are consistent with the numerical results as well as the previous results in the literature, where some approximations were employed.

Our exact formulation of the wave scattering from the KNdS black hole provides simple and practical formulae, which are arbitrary-precision arithmetics.
Unlike known (semi-)analytic calculations in the literature, we do not use any approximations.
There is no restriction on parameters such as the frequency of the scattered waves, or the relative locations of the source of the waves, black hole, and observer.
The exact formulae predict the scattering problem with arbitrary high accuracy.
While we have highlighted several specific applications, it would be intriguing to apply our formulation to more general cases or other observables.
We leave these topics for future work.

\acknowledgements

H.M.\ was supported by Japan Society for the Promotion of Science (JSPS) Grant-in-Aid for Scientific Research (KAKENHI) No.\ JP18K13565. 
S.N.\ gratefully acknowledges the hospitality of Kogakuin University, where part of this work was done, and thanks Yasusada Nambu of Nagoya University for fruitful discussions.

\bibliographystyle{JHEPmod}
\bibliography{ref}

\end{document}